\documentclass[twocolumn,showpacs,preprintnumbers,amsmath,amssymb]{revtex4}
\pdfoutput=1
\usepackage{graphicx}
\usepackage{dcolumn}
\usepackage{bm}
\usepackage{float}

\begin{document}

\title{Dissipative preparation of large W states in Optical Cavities}

\author{Ryan Sweke}
 \email{rsweke@gmail.com}
\affiliation{Quantum Research Group, School of Chemistry and Physics, and National Institute
for Theoretical Physics, University of KwaZulu-Natal, Durban, 4001, South Africa}

\author{Ilya Sinayskiy}
 \email{sinayskiy@ukzn.ac.za}
\affiliation{Quantum Research Group, School of Chemistry and Physics, and National Institute
for Theoretical Physics, University of KwaZulu-Natal, Durban, 4001, South Africa}

\author{Francesco Petruccione}
 \email{petruccione@ukzn.ac.za}
\affiliation{Quantum Research Group, School of Chemistry and Physics, and National Institute
for Theoretical Physics, University of KwaZulu-Natal, Durban, 4001, South Africa}

\date{\today}

\begin{abstract}

Two novel schemes are proposed for the dissipative preparation of large W states, of the order of ten qubits, within the context of Cavity QED. By utilizing properties of the irreducible representations of su(3), we are able to construct protocols in which it is possible to restrict the open system dynamics to a fully symmetric irreducible subspace of the total Hilbert space, and hence obtain analytic solutions for effective ground state dynamics of arbitrary sized ensembles of $\Lambda$ atoms within an optical cavity. In the proposed schemes, the natural decay processes of spontaneous emission and photon loss are no longer undesirable, but essential to the required dynamics. All aspects of the proposed schemes relevant to implementation in currently available optical cavities are explored, especially with respect to increasing system size.

\end{abstract}


\maketitle

\section{INTRODUCTION}

A path towards the experimental realization of a quantum computer has become one of the main focus areas of current research \cite{compreal1}. Many quantum algorithms have been designed and studied \cite{compreal2}-\cite{compreal3}, however in order for their implementation to become a reality it is essential to be capable of creating and manipulating large scale entanglement between effective physical qubits. One of the primary obstacles in this regard is the interaction of a system with its environment, resulting in dissipation and decoherence \cite{franbook}. An effective strategy in combating these destructive effects on unitary implementations of quantum algorithms has been the introduction of error-correcting codes \cite{compreal4}. This approach is based on treating the system-environment interaction as a negative influence, the effect of which needs to be minimized.

A recent paradigm shift in the approach towards the physical realisation of a quantum computer has been introduced by the theoretical prediction that dissipation can in fact be utilized for the creation of complex entangled states \cite{prep0}-\cite{prep6} and to perform universal quantum computation \cite{dcomp1}-\cite{OQW}. This fundamental shift in approach is based on the assumption that the system environment coupling can be manipulated such that the system is driven towards a steady state which is the solution to a computational task, or a desired entangled state \cite{dcomp1}. Within this approach dissipation is no longer a negative effect, but crucial to the required dynamics. Recent experimental progress with atomic ensembles \cite{expreal1} and trapped ions \cite{expreal2}-\cite{expreal3} has shown this approach to be both feasible and promising.

Concurrently many protocols have been suggested for physical dissipative state engineering within cavity QED setups \cite{prep1}-\cite{prepW}. These schemes suggest that it is possible to prepare maximally entangled states of two qubits \cite{prep1}-\cite{prep4}, as well as the maximally entangled W state of three qubits \cite{prepW}, with excellent fidelities, scaling better in cavity cooperativity than any known coherent unitary protocols \cite{prep1}.

Dissipative schemes utilising $\Lambda$ atoms within optical cavities have been particularly successful and well studied \cite{prep1}-\cite{prepW}, however as of yet no scheme has been suggested for which scaling of the scheme to large numbers of atoms is possible. In this work we suggest a physical scheme and a mathematical framework, which in conjunction with the Effective Operator Formalism for adiabatic elimination \cite{effop}, makes it possible to derive an analytic solution for the effective two-level ground state dynamics of arbitrary sized ensembles of $\Lambda$ atoms within an optical cavity. Moreover, we demonstrate the possibility of engineering parameters within a bimodal cavity such that it is possible to prepare large W states, irrespective of the initial thermal state of the system, with excellent fidelities and scaling characteristics. 

We proceed by introducing preliminary theory, before demonstrating the implementation of our suggested method within a single-mode cavity, in which one is restricted to specific initial states of the system. Finally, we present our scheme for the dissipative preparation of large W states, irrespective of the initial thermal state of the system, within a bimodal cavity.

\section{Preliminary Theory}

\begin{figure}
\begin{center}
\includegraphics[width=0.7\linewidth]{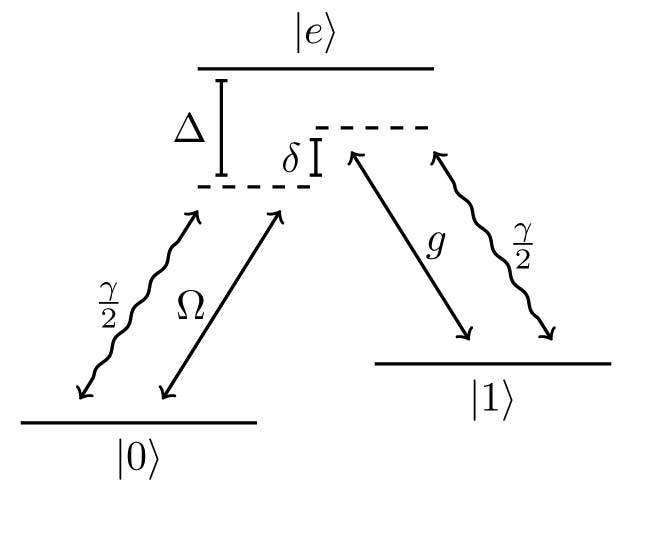} 
\caption{Cavity QED setup for a single atom. The $|0\rangle - |e\rangle$ transition is driven by a coherent laser with a resonant Rabi frequency of $\Omega$ and a detuning of $\Delta$, while levels $|1\rangle$ and $|e\rangle$ are coupled via the cavity field, with an atom-cavity interaction strength of $g$. The entire set up consists of $n$ identical atoms within a single cavity.}\label{threelevel} 
 \end{center}
\end{figure}

We use a cavity QED setup of three-level $\Lambda$ atoms within an optical cavity, as per \cite{prep1}-\cite{prepW}. As per Figures \ref{threelevel} and \ref{bimodalatom}, each $\Lambda$ atom consists of two ground states, $|0\rangle$ and $|1\rangle$, and an excited state $|e\rangle$, coupled to cavity modes. The Hamiltonian for the system is given  by

\begin{equation} \label{hamform}
\hat{H} = \hat{H}_{g} + \hat{H}_{e} + \hat{W}_{+} + \hat{W}_{-},
\end{equation}
where $\hat{H}_{e}$ is the Hamiltonian for the excited subspace, $\hat{H}_{g}$ the Hamiltonian for the ground subspace, $\hat{W}_{+}$ the perturbative excitation from the ground space to the excited space and $\hat{W}_{-}$ the perturbative de-excitation. 

The total system, which consists of a collection of three level $\Lambda$ atoms and a single quantised mode of the cavity electromagnetic field, interacts with an external thermal environment. As is typical for such cavity QED systems in vacuum \cite{prep1}-\cite{prepW} (with experimental realizations described in \cite{new1}-\cite{new3}), the unitary interaction between the ensemble of three-level atoms and the photon mode is damped by both spontaneous emission from the excited states of the three level atoms and decay of the photon mode. The dynamics of the system is described by a master equation, incorporating the Born-Markov approximation, in GKSL form \cite{franbook}.

\begin{align}\label{masterequation}
 \dot{\rho} &= \mathcal{L}\rho = -i[\hat{H},\rho]\nonumber  \\&+ \sum_{k}\left(\hat{L}_{k}\rho \hat{L}^{\dagger}_{k} - \frac{1}{2}\hat{L}_{k}\hat{L}^{\dagger}_{k}\rho - \frac{1}{2}
\rho \hat{L}^{\dagger}_{K}\hat{L}_{k}\right).
\end{align}

All previously suggested schemes \cite{prep1}-\cite{prepW} require non-uniform individual laser addressing of atoms within the cavity. This requirement makes realistic scaling and generalization to larger atomic ensembles impossible, and motivates the use of protocols designed around global uniform addressing of atoms within the cavity. Mathematically this corresponds to a Hamiltonian and Lindblad operators constructed from collective operators of the form

\begin{equation}
\hat{O} = \sum_{i = 1}^{n} \hat{O}_{i},
\end{equation}
where for systems of $n$ $\Lambda$-atoms, $\hat{O}_{i}$ acts on the states within the Hilbert space of the $i$'th atom, and the total Hilbert space for the atomic ensemble is the direct product of the $n$ individual Hilbert spaces. In analogy with methods for the solution of arbitrary sized ensembles within the Dicke Model \cite{dicke1}-\cite{dicke4}, we will show that if one devises a physical system in which the Hamiltonian and Lindblad operators are formed from specific collective operators, then specific subspaces of the total Hilbert space are invariant under the action of both the Hamiltonian and Lindblad operators. In the case of the Dicke model these irreducible subspaces, invariant under the action of Hamiltonian and Lindblad operators formed from collective generators of SU(2), are the irreducible representations of su(2). However for arbitrary sized ensembles of $\Lambda$ atoms it is natural to examine the irreducible representations of su(3).

Furthermore, the Effective Operator Formalism \cite{effop} has provided an extremely elegant method for performing adiabatic elimination \cite{adelim1}, such that it is possible to isolate effective ground state dynamics. For  an optical cavity QED setup with Hamiltonian of the form \eqref{hamform}, described by a master equation as in Eq. \eqref{masterequation}, one can obtain an effective master equation \cite{effop},

\begin{align}\label{effmastereq}
 \dot{\rho}_{g} &= -i[\hat{H}_{eff},\rho_{g}]  + \sum_{k}\Big(\hat{L}^{k}_{eff}\rho_{g} (\hat{L}^{k}_{eff})^{\dagger}  \nonumber\\
 & -  \frac{1}{2}\hat{L}^{k}_{eff}(\hat{L}^{k}_{eff})^{\dagger}\rho_{g} - \frac{1}{2}
\rho_{g} (\hat{L}^{k}_{eff})^{\dagger}\hat{L}^{k}_{eff} \Big),
\end{align}
where $\rho_{g}$ is the density matrix for the ground subspace and

\begin{align}
&\hat{H}_{eff} \equiv- \frac{1}{2}\hat{W}_{-}\Big( \hat{H}_{NH}^{-1} + (\hat{H}_{NH}^{-1})^{\dagger}\Big)\hat{W}_{+} + \hat{H}_{g},  \label{heffdef}\\
& \hat{L}^{k}_{eff} \equiv \hat{L}_{k} \hat{H}_{NH}^{-1} \hat{W}_{+}, \label{leffdef}
\end{align}
with,

\begin{equation}\label{HNHdef}
\hat{H}_{NH} \equiv \hat{H}_{e} - \frac{i}{2}\sum_{k}\hat{L}_{k}^{\dagger}\hat{L}_{k}.
\end{equation}

The use of adiabatic elimination implies a restriction to the single excitation subspace of the atom-cavity system, for which certain physical assumptions are necessary, which are discussed in Section III. As we would like to isolate effective ground state dynamics this restriction, in conjunction with a consideration of the consequences of a collective operator approach, motivates an investigation of the single excitation irreducible subspaces of the total atom-cavity Hilbert space, especially with respect to their invariance under specific collective operators. In order to construct these invariant subspaces, and determine their irreducibility properties, we proceed via analogy with the Dicke model.

A familiar single spin-half system, as in the Dicke Model, exists within a Hilbert space whose basis consists of the two eigenvectors of $S_{z}$, denoted here by the kets $|0\rangle, |1\rangle$. These two kets form a multiplet which can be considered the fundamental representation of su(2), the angular momentum Lie algebra and generator of the symmetry group SU(2). For a system of multiple spin-half particles, as per the theory for the addition of angular momenta, the total Hilbert space consists of multiple invariant irreducible subspaces, spanned by multiplets which are irreducible representations of su(2). For example, it is well known that the total Hilbert space of a system consisting of two spin-half particles consists of an invariant symmetric subspace, spanned by the triplet multiplet, and an invariant antisymmetric subspace, spanned by the singlet state. Mathematically, constructing these multiplets requires a reduction of the Hilbert space from a tensor product of two spin-half spaces, into the direct sum of a spin one Hilbert space, and a spin zero Hilbert space, a process generally described by using the following notation,

\begin{equation}
\bigg[\frac{1}{2}\bigg]\otimes\bigg[\frac{1}{2}\bigg] = [1]\oplus[0].
\end{equation}
These subspaces of the total Hilbert space are invariant in the sense that they are closed under the action of elements of SU(2), and irreducible in the standard sense that they contain no smaller invariant subspaces.
 
 As we are dealing with a collection of $\Lambda$ atoms, the total atomic Hilbert space is given by
 
 \begin{equation}
\mathcal{H} = \bigotimes_{i = 1}^{n} \mathbb{C}^3 
\end{equation}
and hence it is natural to examine the invariant irreducible subspaces of su(3), the underlying symmetry group relevant to this problem. Following \cite{sym} we construct these invariant irreducible subspaces of the total atomic Hilbert space through a generalized angular momentum approach. This process will involve the reduction of the direct product of individual atomic Hilbert spaces, each spanned by a fundamental representation of su(3), into a unique direct sum of subspaces spanned by irreducible representations of su(3). 

This approach is in direct analogy to the approach taken in particle physics, where the irreducible representations of su(3) are used to construct Baryon and Meson multiplets from up, down and strange quarks and antiquarks, which collectively form the two fundamental representations of su(3) in that context \cite{sym}.

In general, SU(n) has $n^2 - 1$ generators, and hence SU(3) has 8. These generators are typically denoted as

\begin{equation}
\hat{\lambda}_{1},\hat{\lambda}_{2},\ldots,\hat{\lambda}_{8}.
\end{equation}
From the theory of Lie Groups, each $\hat{\lambda}_{i}$ is required to be both Hermitian and traceless. As a consequence of the fact that SU(2) $ \subset $ SU(3), the first three generators are constructed by extending the familiar generators of SU(2) into 3 dimensions. The rest of the generators can then be chosen in a variety of manners. We choose to follow the conventions of particle physics. The full matrix form of all the generators can be found in the appendix.

From the full form of the generators $\{\hat{\lambda}_{i}\}$, one can calculate their commutators, and hence the Lie algebra su(3). These commutation relations are found to be

\begin{equation}
\label{commute}
[\hat{\lambda}_{i},\hat{\lambda}_{j}] = 2 i f_{ijk}\hat{\lambda}_{k},
\end{equation}
where the structure constants are totally antisymmetric under the exchange of any two indices, and can be found in detail in the appendix. From Eq. \eqref{commute} it can be seen that, as expected, the Lie Algebra of SU(3) is indeed closed. In analogy with angular momentum in SU(2) it is helpful to redefine the generators as

\begin{equation}
\hat{F}_{i} = \frac{1}{2}\hat{\lambda}_{i}.
\end{equation}
From \eqref{commute} it then follows that

\begin{equation}\label{commute2}
[\hat{F}_{i},\hat{F}_{j}] =  i f_{ijk}\hat{F}_{k}.
\end{equation}
In particle physics, the clear analogy between the above, and the familiar angular momentum situation in SU(2), has led to generators $\{\hat{F}_{i}\}$ being labelled as F-spin. In order to continue with this generalization, we proceed to introduce the following representation of the F-spin operators,

\begin{equation}\label{Fspin1}
\hat{T}_{\pm} = \hat{F}_{1} \pm i\hat{F}_{2}, \qquad \hat{T}_{3} = \hat{F}_{3}, 
\end{equation}
\begin{equation}\label{Fspin2}
\hat{V}_{\pm} = \hat{F}_{4} \pm i\hat{F}_{5}, \qquad \hat{U}_{\pm} = \hat{F}_{6} \pm i\hat{F}_{7},
\end{equation}
\begin{equation}\label{Fspin3}
\hat{Y} =\frac{2}{\sqrt{3}} \hat{F}_{8}. 
\end{equation}
The full set of commutation relations for the above operators is of great importance to what follows, and can be found in the appendix.

The structure of the irreducible representations of su(3) follows from the existence of sub-algebras. In order to see this we note that the commutation relationships,

\begin{equation}
[\hat{T}_{+},\hat{T}_{-}] = 2\hat{T}_{3} \quad [\hat{T}_{3},\hat{T}_{\pm}] = \pm\hat{T}_{\pm},
\end{equation}
show that the operators $\{\hat{F}_{1},\hat{F}_{2},\hat{F}_{3}\}$ fulfill the algebra of su(2), and hence the operators $\{\hat{T}_{+},\hat{T}_{-},\hat{T}_{3}\}$ form a closed sub-algebra of su(3). Similarly, we have that

\begin{equation}
[\hat{U}_{+},\hat{U}_{-}] = 2\hat{U}_{3}, \quad [\hat{U}_{3},\hat{U}_{\pm}] = \pm\hat{U}_{\pm},
\end{equation}
\begin{equation}
[\hat{V}_{+},\hat{V}_{-}] = 2\hat{V}_{3}, \quad [\hat{V}_{3},\hat{V}_{\pm}] = \pm\hat{V}_{\pm},
\end{equation}
and hence the operator sets  $\{\hat{U}_{+},\hat{U}_{-},\hat{U}_{3}\}$ and  $\{\hat{V}_{+},\hat{V}_{-},\hat{V}_{3}\}$ both also form closed sub-algebras of su(3), where $\hat{U}_{3}$ and $\hat{V}_{3}$ are still to be defined. All three of these closed sub-algebras  match the algebra of the familiar angular momentum operators. The action of these operators is made clear by considering the commutation relationship,

\begin{equation}
[\hat{Y},\hat{T}_{3}] = 0,
\end{equation}
which shows that the operators $\hat{Y}$ and $\hat{T}_{3}$ may be simultaneously diagonalized. If we represent their common eigenstate by $|\hat{T}_{3},\hat{Y}\rangle$, then it follows that

\begin{equation}
\hat{T}_{3}|\hat{T}_{3},\hat{Y}\rangle = T_{3}|\hat{T}_{3},\hat{Y}\rangle, 
\end{equation}
\begin{equation}
\hat{Y}|\hat{T}_{3},\hat{Y}\rangle = Y|\hat{T}_{3},\hat{Y}\rangle,
\end{equation}
from which it is possible to show that

\begin{equation}
\hat{T}_{3}(\hat{V}_{\pm}|\hat{T}_{3},\hat{Y}\rangle) = (T_{3}\pm\frac{1}{2})(\hat{V}_{\pm}|\hat{T}_{3},\hat{Y}\rangle),
\end{equation}
which implies that $\hat{V}_{\pm}$ transforms a state with quantum number $T_{3}$ into a state with quantum number $T_{3} \pm \frac{1}{2}$. Similarly, it can be shown that

\begin{equation}
\hat{T}_{3}(\hat{U}_{\pm}|\hat{T}_{3},\hat{Y}\rangle) = (T_{3}\mp\frac{1}{2})(\hat{U}_{\pm}|\hat{T}_{3},\hat{Y}\rangle),
\end{equation}
and hence $\hat{U}_{\pm}$ lowers and raises, respectively, the quantum number $T_{3}$ by $\frac{1}{2}$. It is also clear, by construction and from analogy with angular momentum, that  $\hat{T}_{\pm}$ raises and lowers the quantum number $T_{3}$ by integer units.

\begin{figure} 
\includegraphics[width=0.7\linewidth]{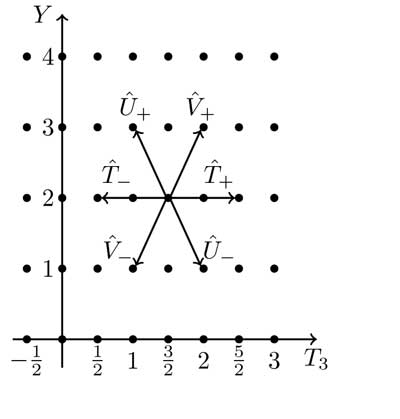} 
 \caption{Action of shift operators in the $(T_3, Y)$ plane. The units are scaled such that one unit on the $Y$ axis is $\sqrt{3/4}$ times a single unit on the $T_3$ axis.} \label{operatoraction}
\end{figure}

From the commutators,

\begin{equation}
[\hat{Y},\hat{V}_{\pm}] = \pm\hat{V}_{\pm}, \quad [\hat{Y},\hat{U}_{\pm}] = \pm\hat{U}_{\pm},
\end{equation}
it can be shown that 

\begin{equation}
\hat{Y}(\hat{U}_{\pm}|\hat{T}_{3},\hat{Y}\rangle) = (Y\pm 1)(\hat{U}_{\pm}|\hat{T}_{3},\hat{Y}\rangle),
\end{equation}
\begin{equation}
\hat{Y}(\hat{V}_{\pm}|\hat{T}_{3},\hat{Y}\rangle) = (Y\pm 1)(\hat{V}_{\pm}|\hat{T}_{3},\hat{Y}\rangle),
\end{equation}
and hence $\hat{V}_{\pm}$ and $\hat{U}_{\pm}$ both raise and lower, respectively, the quantum number $Y$ by integer units. Finally, from the commutator $[\hat{Y},\hat{T}_{\pm}] = 0$ it is possible to see that the operators $\hat{T}_{\pm}$ do not change the value of the $Y$ quantum number. The action of all these operators is shown in Figure \ref{operatoraction}.

Armed with the above it is possible to gain an insight into the structure of SU(3) multiplets, the irreducible representations of su(3). As the $T$, $U$ and $V$ algebra's, all isomorphic to the algebra of angular momentum, all form sub-algebras of SU(3), the SU(3) multiplets can be constructed from coupled $T$, $U$ and $V$ multiplets. Figure \ref{operatoraction} illustrates the fact that the $T$ multiplets lie parallel to the $T_{3}$ axis, the $V$ multiplets lie along $V$ lines, and the $U$ multiplets lie along $U$ lines. Commutation relationships such as $[\hat{T}_{+},\hat{V}_{-}] = -\hat{U}_{-}$ and $[\hat{T}_{+},\hat{U}_{+}] = -\hat{V}_{+}$ force the coupling of these  SU(2) sub-multiplets to form SU(3) multiplets.

The structure of the SU(3) multiplets in the $(Y,T_{3})$ plane follows from considering the structure of the individual SU(2) sub-multiplets, and the relationships between the three sub-algebras. From the theory of angular momentum in SU(2) it follows that the $T_{3}$ values of all members of the $T$-algebra sub-multiplet are within the interval $T_{3 (\mathrm{min})}\leq T_{3} \leq T_{3 (\mathrm{max})}$ and hence the $T$-algebra sub-multiplet is symmetric around the $T_{3} = 0$ axis. As the  $T$, $U$ and $V$ sub-algebras are completely equivalent, and hence equally symmetric, the $U$-algebra sub-multiplet will be symmetric around the $2U_{3} = \frac{3}{2}Y - T_{3} = 0$ axis and the $V$-algebra sub-multiplet will be symmetric around the $2V_{3} = \frac{3}{2}Y + T_{3} = 0$ axis. The SU(3) multiplets formed from coupling $T-$, $U-$ and $V-$ algebra sub-multiplets will therefore be symmetric with respect to the $T_{3} = 0$, $U_{3} = 0$ and $V_{3} = 0$ axis, resulting in SU(3) multiplets which are either regular hexagonal, or triangular in the $(Y,T_{3})$ plane. It also follows, by construction, that all SU(3) multiplets will be centred around the origin of the $(Y,T_{3})$ plane, and invariant under rotations of $2\pi/3$ about the origin.

It is now  necessary to consider the structure of SU(3) multiplets in more detail. Every mulitplet will contain one state, described as the state with \emph{maximal weight}, and denoted $|\psi_M\rangle$, associated with the largest $T_{3}$ value in the multiplet. From Figure \ref{operatoraction} it is clear that this state has the property

\begin{equation}
\hat{T}_{+}|\psi_M\rangle = \hat{V}_{+}|\psi_M\rangle = \hat{U}_{-}|\psi_M\rangle = 0.
\end{equation} 
If one can identify this state, then the boundary of the multiplet can be constructed in the following algorithmic manner. From $|\psi_M\rangle$, successive member-states of the boundary can be achieved by repeated application of $\hat{V}_{-}$. After $p$ applications of $\hat{V}_{-}$ a state will be reached such that

\begin{equation}
(\hat{V}_{-})^{p+1}|\psi_M\rangle = 0,
\end{equation}
uniquely defining the integer $p$. From the state $(\hat{V}_{-})^{p}|\psi_M\rangle$ the boundary of the multiplet can be continued by successive applications of  $\hat{T}_{-}$,  until after $q$ applications one reaches the state such that

\begin{equation}
(\hat{T}_{-})^{q+1}(\hat{V}_{-})^{p}|\psi_M\rangle = 0,
\end{equation}

uniquely defining the integer $q$. The two integers $(p,q)$ define  SU(3) multiplets, as the remaining boundary states follow necessarily from considerations of symmetry discussed above.

  As discussed in detail in \cite{sym}, for a given multiplet $(p,q)$, the states on the boundary of the multiplet are unique (i.e., each mesh point on the hexagonal boundary of the multiplet corresponds to only one state), however as one moves through inner hexagonal shells of the multiplet, the multiplicity of each mesh point (the number of different states associated with that point) increases by 1 with each step towards the origin, until after $q$ steps (where $q \leq p$), the hexagon has become a triangle, and the multiplicity of each mesh point is $q+1$.

At this point we have sufficient information to construct the invariant irreducible subspaces of the total Hilbert space of an arbitrary number of $\Lambda$-atoms, or in the language of particle physics, to decompose the direct product of $n$ Hilbert spaces into the direct sum of irreducible invariant subspaces. For a system of $n$ $\Lambda$ atoms, notice that the collective operators

\begin{equation}\label{operator1}
\hat{T}_{+} = \sum_{i=1}^{n}|0\rangle_{i}\langle 1|, \qquad \hat{T}_{-} = \sum_{i=1}^{n}|1\rangle_{i}\langle 0|,
\end{equation}
\begin{equation}\label{operator2}
\hat{V}_{+} =\sum_{i=1}^{n} |0\rangle_{i}\langle e|, \qquad \hat{V}_{-} =\sum_{i=1}^{n} |e\rangle_{i}\langle 0|,
\end{equation}
\begin{equation}\label{operator3}
\hat{U}_{+} = \sum_{i=1}^{n}|1\rangle_{i}\langle e|, \qquad \hat{U}_{-} = \sum_{i=1}^{n}|e\rangle_{i}\langle 1|,
\end{equation}
\begin{equation}\label{operator4}
\hat{T}_{3} = \frac{1}{2}\sum_{i=1}^{n}\bigg(|0\rangle_{i}\langle 0| - |1\rangle_{i}\langle 1| \bigg),
\end{equation}
\begin{equation}\label{operator5}
\hat{Y} = \frac{1}{2\sqrt{3}} \sum_{i=1}^{n}\bigg(|0\rangle_{i}\langle 0| + |1\rangle_{i}\langle 1| - 2|e\rangle_{i}\langle e| \bigg),
\end{equation}
fulfil all the commutation relationships (\ref{commute1})-(\ref{commute7}), and hence are suitable representations of generators for SU(3). Utilizing operators \eqref{operator1} - \eqref{operator5} it is now possible to apply the discussed theory in order to create multiplets of states, each of which spans a unique invariant irreducible subspace of the total Hilbert space. 

\begin{figure} 
 \includegraphics[width=0.7\linewidth]{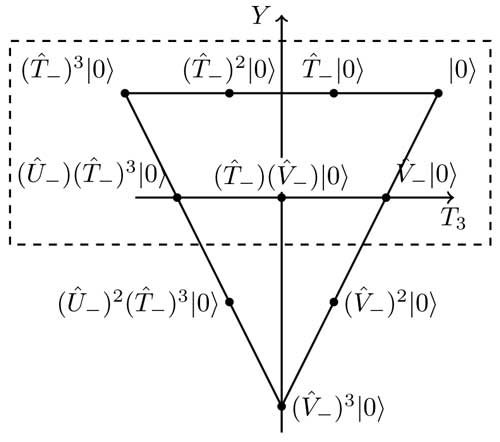} 
 \caption{Construction of the multiplet $(3,0)$ in the $(T_3,Y)$ plane. The top row of the multiplet consists of ground states, while the second row consists of single atomic-excitation states.} \label{multipletthree}
\end{figure}

However, as the subspace spanned by each possible multiplet is invariant under the action of collective operators \eqref{operator1} - \eqref{operator5}, it is clear that if one constructs a Hamiltonian from the above collective operators, then it is sufficient to restrict one's analysis to the invariant subspace containing the desired target state of our scheme, the W state. As the W state is a symmetric state, this implies that it is only necessary to construct the completely symmetric multiplets, spanning completely symmetric subspaces, provided we construct the Hamiltonian of our scheme from collective operators \eqref{operator1} - \eqref{operator5}. It is useful to begin with an analysis of a system of 3 $\Lambda$ atoms.

In this case the symmetric state with maximal weight, $|\psi_{(M,S)}\rangle$, is

\begin{equation}
|\psi_{(M,S)}\rangle = |000\rangle.
\end{equation}
It is important to note that the W state that we are interested in creating, which is the state containing the maximum amount of sum of two qubit entanglement \cite{Wstate}, indeed belongs to this multiplet and is given by

\begin{align}
|W\rangle &= \frac{1}{\sqrt{3}}(|001\rangle + |010\rangle + |100\rangle) \\
&= \Big(\frac{1}{\sqrt{3}}\Big) \hat{T}_{-}|\psi_{(M,S)}\rangle.
\end{align}
In order to determine the properties of this multiplet, we note that

\begin{equation}
(\hat{V}_{-})^{(3+1)}|\psi_{(M,S)}\rangle = 0,
\end{equation}

and, 

\begin{equation}
(\hat{T}_{-})^{(0+1)}(\hat{V}_{-})^{3}|\psi_{(M,S)}\rangle = 0.
\end{equation}
Hence, the symmetric multiplet of 3 $\Lambda$ atoms is the unique multiplet $(p,q) = (3,0)$. As per \cite{sym}, the number of states within a multiplet, $d(p,q)$, is given by

\begin{equation}
d(p,q) = \frac{1}{2}(p+1)(q+1)(p+q+2),
\end{equation}

and there are 10 states in the multiplet $(3,0)$ under consideration. This multiplet is constructed in the manner previously described, and is shown in Figure \ref{multipletthree}.

From Figure \ref{multipletthree} one can see that the top row of the multiplet consists only of ground states, while the second row consists of states with a single excitation and the remaining rows contain states with more than a single excitation (as per the conventional interpretation of a $\Lambda$ atom). Therefore, as adiabatic elimination implies a restriction to the single excitation subspace, we are only concerned with the top two rows of the multiplet, which form a basis for the completely symmetric, single excitation subspace of three $\Lambda$ atoms.

\begin{figure*}[t] 
\includegraphics[width=0.7\linewidth]{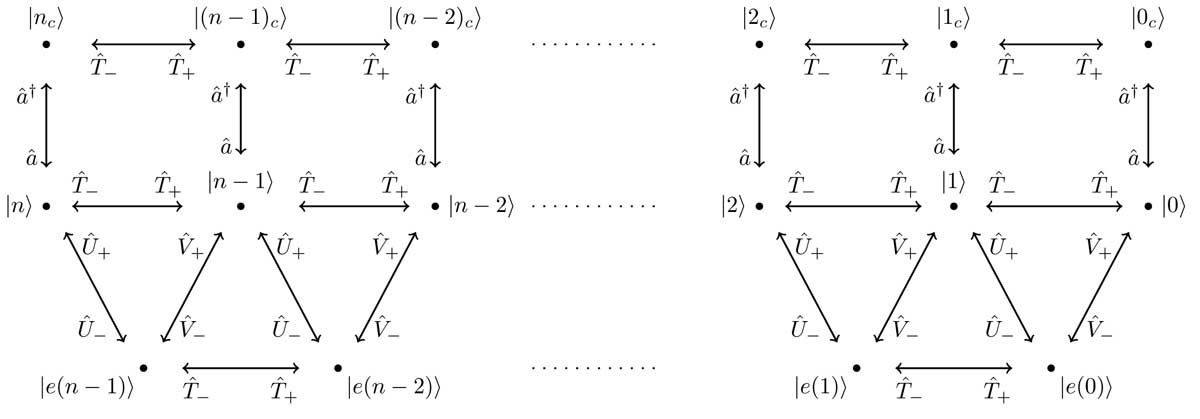}
\caption{Construction of the total basis for the symmetric, single-excitation subspace of $n$ $\Lambda$ atoms in a single optical cavity. The top row consists of single cavity-excitation states, the second row of ground states and the third row of single atomic-excitation states. The notation is as per Eqs. (\ref{not1} - \ref{not3}).  }\label{multipletarb} 
\end{figure*}

It is now possible to generalize this to the case of $n$ atoms. The symmetric state with maximal weight is given by

\begin{equation}
|\psi_{(M,S)}\rangle = \bigotimes_{i=1}^{n}|0\rangle_{i}
\end{equation}

and in general,

\begin{equation}
(\hat{V}_{-})^{(n+1)}|\psi_{(M,S)}\rangle = 0,
\end{equation}

and

\begin{equation}
(\hat{T}_{-})^{(0+1)}(\hat{V}_{-})^{n}|\psi_{(M,S)}\rangle = 0,
\end{equation}
such that the completely symmetric multiplet for $n$ atoms, the basis for the completely symmetric subspace, is the multiplet $(p,q) = (n,0)$. This multiplet is again triangular and only the first two rows contain states from the single excitation subspace. The number of these single excitation symmetric states, $d_{1}(p)$, is given by

\begin{equation}
d_{1}(p) = (p+1) + p = 2p + 1.
\end{equation}
It can now be seen that a basis for the completely symmetric, first excitation subspace of $n$ $\Lambda$ atoms, is given by the union of the following two sets,

\begin{equation}\label{G}
G = \left\{ |0\rangle, \ldots ,  \frac{1}{\sqrt{ \binom{n}{j} } }|j\rangle , \ldots, |n\rangle  \right\},
\end{equation}

\begin{equation}\label{A}
A = \left\{ \frac{1}{\sqrt{n}}|e (0) \rangle, \ldots ,  \frac{1}{\sqrt{ n \binom{n-1}{j} } }|e (j)\rangle , \ldots, \frac{1}{\sqrt{n}}|e(n-1)\rangle  \right\},
\end{equation}

where

\begin{equation}\label{not1}
|j\rangle = (\hat{T}_{-})^{j}|\psi_{(M,S)}\rangle,
\end{equation}

and

\begin{equation}
|e (j)\rangle = (\hat{T}_{-})^{j}(\hat{V}_{-})|\psi_{(M,S)}\rangle.
\end{equation}
In this case, $G$ is the set of completely symmetric ground states, the top row of the $(n,0)$ multiplet in the $(T_{3},Y)$ plane, and $A$ is the set of completely symmetric, single atomic excitation states, the second from the top row of the $(n,0)$ multiplet.

While $\{G,A\}$ forms a basis for the completely symmetric, first excitation subspace of $n$ $\Lambda$ atoms, we are actually interested in $n$ $\Lambda$ atoms within a single optical cavity, and later within a single bimodal optical cavity. The Hilbert space for an empty single-mode cavity, restricted to one excitation, is $\mathbb{C}^2$, with a basis we choose to denote $\{|0_{c}\rangle,|1_{c}\rangle\}$. If we now adopt the notation,

\begin{equation} \label{not3}
|j\rangle = |j\rangle\otimes|0_c\rangle \quad |j_{c}\rangle = |j\rangle\otimes|1_c\rangle,
\end{equation}
then a basis for the  symmetric, single-excitation subspace of $n$ $\Lambda$ atoms within a single optical cavity, is a union of the sets $G$, $A$ and $C$ where

\begin{equation}
C = \left\{ |0_{c}\rangle, \ldots ,  \frac{1}{\sqrt{ \binom{n}{j} } }|j_{c}\rangle , \ldots, |n_{c}\rangle  \right\}.
\end{equation}
It is important to note that $C$ is the set of completely symmetric, single cavity-excitation states, created naturally from members of the set $G$ by application of the creation operator,

\begin{equation}
\hat{a}^{\dagger} = |1_{c}\rangle\langle 0_{c}|.
\end{equation}
As $C$ has the same dimension, $(n+1)$, as $G$, the dimension of the full basis for the completely symmetric, single-excitation subspace of $n$ $\Lambda$ atoms within a single optical cavity, $\{G,A,C\}$, is given by

\begin{equation}
d(n) = 3n + 2.
\end{equation}
Figure \ref{multipletarb} illustrates the construction of this basis in full detail.

\section{Dissipative State Preparation}

\subsection{Single Mode Cavity scheme}

In this section we present a simple scheme for the dissipative preparation of large W-states, under the assumption that the system begins in a specified ground state. By deliberate construction of the Hamiltonian from generators of SU(3), we are able to utilize the symmetric, single-excitation basis previously constructed in order to be able to apply the effective operator formalism of \cite{effop} to arbitrary size systems. This allows us to obtain effective operators whose strengths can be engineered, via suitable parameter choices, such that the target state is prepared efficiently and reliably.

We use a cavity QED setup of $n$ $\Lambda$ atoms in a single-mode optical cavity. As per Figure \ref{threelevel}, each $\Lambda$ atom consists of two ground states, $|0\rangle$ and $|1\rangle$, and an excited state $|e\rangle$, which is coupled to a cavity mode. The Hamiltonian for the system is of the form given in Eq. \eqref{hamform}, where in the appropriate rotating frame the Hamiltonian is time independent with the following individual terms,

\begin{align}
&\hat{H}_{e} = \Delta \sum_{i=1}^{n}|e\rangle_{i}\langle e | + \delta(\hat{a}^{\dagger}\hat{a}) + \hat{H}_{ac},\\
&\hat{H}_{ac} = g \Big( \hat{a}^{\dagger}\hat{U}_{+} + h.c \Big),\\
&\hat{W}_{+} =\frac{\Omega}{2}\hat{V}_{-},\\
&\hat{W}_{-} =\hat{W}_{+}^{\dagger}, \\
&\hat{H}_{g} = 0,
\end{align}
where $\hat{V}_{-}$ and $\hat{U}_{+}$ are as per Eqs. \eqref{operator2} and \eqref{operator3}. The perturbative excitation, $\hat{W}_{+}$, is driven by a coherent laser, addressing all atoms uniformly, with a resonant Rabi frequency $\Omega$ and a detuning of $\Delta$, while the atom-cavity interaction term, $\hat{H}_{ac}$, describes the coupling of levels $|e\rangle$ and $|1\rangle$ via the cavity field, with a strength of $g$ and uniform phase over all atoms. It is important to note that the atom-cavity coupling for each atom depends on the cavity mode functions and is therefore not \emph{a priori} the same for each atom. However, in currently available optical cavities \cite{kimbleQED}, mirror sizes and cavity scales are such that, for the number of atoms relevant to this paper, uniform atom-cavity couplings can be obtained through appropriate symmetric arrangement of atoms within the cavity.

As we assume Markovian interaction with the environment, an extremely good assumption within quantum optics \cite{carQO}, as is relevant to this paper, the evolution of the system is described by a master equation of the form given in Eq. \eqref{masterequation}. The Lindblad operator associated with cavity loss, $\hat{L}_{\kappa}$, is given by

\begin{equation}
\hat{L}_{\kappa} = \sqrt{\kappa}\hat{a},
\end{equation}
where $\kappa$ is the photon decay rate. The Lindblad operators associated with spontaneous emission are given by

\begin{align}
&\hat{L}_{(\gamma,0)} = \sqrt{\frac{\gamma}{2}}\hat{V}_{+} = \sqrt{\frac{\gamma}{2}}\sum_{i = 1}^n|0\rangle_{i}\langle e |, \\
&\hat{L}_{(\gamma,1)} = \sqrt{\frac{\gamma}{2}}\hat{U}_{+}=\sqrt{\frac{\gamma}{2}} \sum_{i = 1}^n|1\rangle_{i}\langle e |, 
\end{align}
where the decay rates into states $|0\rangle$ and $|1\rangle$ have been made equal ($\sqrt{\gamma/2}$) for simplicity, while the individual atomic emission Lindblad operators have been collected into collective operators, a natural way to treat the system.

From the construction of the Hamiltonian, and the structure of the symmetric single-excitation subspace detailed in Figure \ref{multipletarb}, it is clear that the symmetric single-excitation subspace is closed under the action of the Hamiltonian and Lindblad operators. Hence, as desired, if the initial state of the system is some state within the symmetric single-excitation subspace, we can restrict our attention to this particular subspace.

We will proceed to use adiabatic elimination, via the effective operator formalism of \cite{effop}, in order to reduce the evolution of the system to effective secondary processes between ground states, described by an effective master equation of the form given in equation \eqref{effmastereq}. We will work within the high cooperativity regime $g^2 \gtrsim \kappa\gamma$ and in addition, in order to apply adiabatic elimination (and motivate a restriction to the single-excitation subspace), it is required that we restrict ourselves to the regime of weak driving $\Omega \ll (g,\kappa,\gamma)$ and simultaneously ensure that the excited energy energy levels are largely detuned from the ground levels, i.e., that $\Delta$ (the detuning of the coherent interaction between $|0\rangle$ and $|e\rangle$) and $\Delta - \delta$ (the detuning of the atom-cavity interaction between $|1\rangle$ and $|e\rangle$) are both large, implying $(\Delta,\Delta - \delta) \sim g$.

\begin{figure}[H]
\begin{center}
\includegraphics[width=0.7\linewidth]{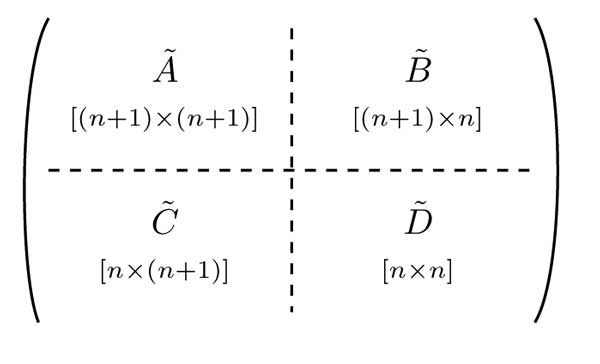}
 \caption{Partitioned matrix form of $\hat{H}_{NH}$}\label{BM1}
 \end{center}
\end{figure}

In the basis of the symmetric first-excitation subspace, and the notation of Eqs. \eqref{not1} - \eqref{not3}, the terms of the Hamiltonian take the following form,

\begin{align}
&\hat{H}_{e} = \Delta \sum_{i=0}^{n-1}|e(i)\rangle\langle e (i)| + \delta\sum_{i=0}^n|i_{c}\rangle\langle i_{c}| + \hat{H}_{ac},\\
&\hat{H}_{ac} = g \Big(\sum_{i=0}^{n-1}(\sqrt{i+1})|(i+1)_{c}\rangle\langle e (i) | + h.c \Big), \\
&\hat{W}_{+} = \frac{\Omega}{2}\sum_{i=0}^{n-1}(\sqrt{n-i})|e(i)\rangle\langle i |,\\
&\hat{W}_{-} = \hat{W}_{+}^{\dagger}, \\
&\hat{H}_{g} = 0,
\end{align}
while the Lindblad operators become,

\begin{align}
& \hat{L}_{1} = \hat{L}_{\kappa} = \sqrt{\kappa}\sum_{i = 0}^{n} |i\rangle\langle i_{c}|, \\
& \hat{L}_{2} = \hat{L}_{(\gamma,0)} = \sqrt{\frac{\gamma}{2}}\sum_{i = 0}^{n-1} |i\rangle\langle e(i)|, \\
& \hat{L}_{3} = \hat{L}_{(\gamma,1)} = \sqrt{\frac{\gamma}{2}}\sum_{i = 0}^{n-1} |i+1\rangle\langle e(i)|.
\end{align}
In this basis it is possible to obtain the effective operators, and the effective Hamiltonian, for arbitrary sized systems. $\hat{H}_{NH}$, the matrix whose inverse consists of propagators between excited states, can be represented as a block matrix with form as per Figure \ref{BM1}, where $\tilde{A}$ is the block pertaining to interactions within the single-cavity excitation subspace, $\tilde{D}$ is the block pertaining to interactions within the single atomic-excitation subspace and $\tilde{B}$,$\tilde{C}$ are blocks describing interactions between the two single-excitation subspaces. 

\begin{figure*}[t] 
\includegraphics[width=0.8\linewidth]{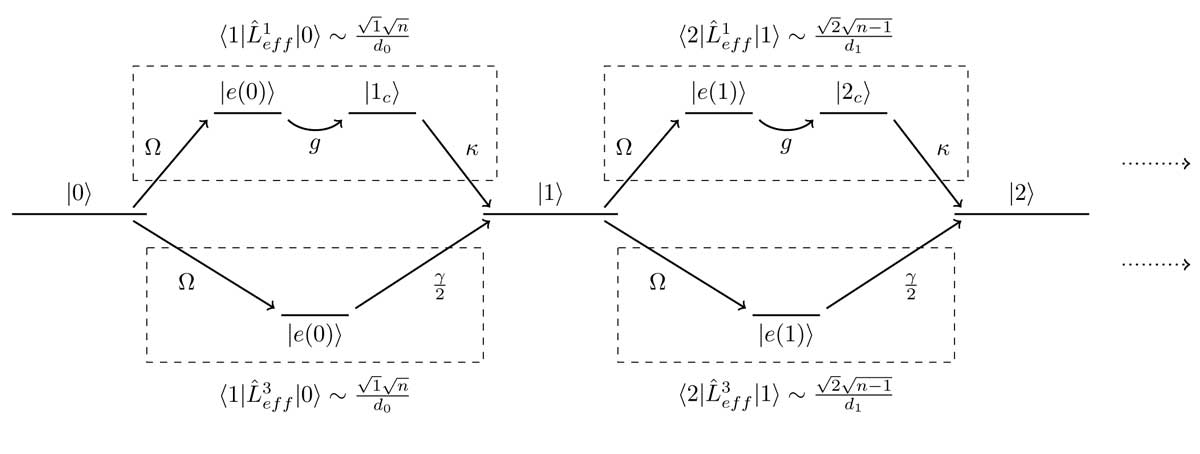} 
\caption{Summary of effective ground state processes, Eqs. \eqref{eff1} - \eqref{effham}, omitting effective ``loop" processes of $\hat{L}^{2}_{eff}$ and $\hat{H}_{eff}$. Each effective process consists of a coherent excitation, an intermediate propagation within the single-excitation subspace (facilitated by $\hat{H}^{-1}_{NH}$) and a de-excitation via coherent driving or dissipation. }\label{effsys} 
\end{figure*}

In the  symmetric single-excitation basis $\hat{H}_{NH}$ is then given by

\begin{equation}
\hat{H}_{NH} = \tilde{A} + \tilde{B} +\tilde{C} + \tilde{D},
\end{equation}
where

\begin{align}
&\tilde{A} = \Big(\delta - i\frac{\kappa}{2}\Big)\sum_{i=0}^n|i_{c}\rangle\langle i_{c}|, \\
&\tilde{D} = \Big( \Delta - i\gamma\frac{(n+1)}{4}\Big)\sum_{i=0}^{n-1}|e(i)\rangle\langle e (i)|, \\
&\tilde{B} = g \Big(\sum_{i=0}^{n-1}(\sqrt{i+1})|(i+1)_{c}\rangle\langle e (i) |, \\
&\tilde{C} = \tilde{B}^{\dagger}.
\end{align}
Using the Banachiewicz inversion theorem for partitioned matrices \cite{matrixinv} we find that $\hat{H}_{NH}^{-1}$, the propagator representing the non-Hermitian evolution of the excited subspace, is given by

\begin{equation}\label{invert1}
\hat{H}_{NH}^{-1} = \hat{A} + \hat{B} + \hat{C} + \hat{D},
\end{equation}
where

\begin{align}
&\hat{D} = \big(\tilde{D} - \tilde{C}\tilde{A}^{-1}\tilde{B}\big)^{-1},   \\
&\hat{A} = \tilde{A}^{-1} + \tilde{A}^{-1}\tilde{B}\big(\tilde{D} - \tilde{C}\tilde{A}^{-1}\tilde{B}\big)^{-1}\tilde{C}\tilde{A}^{-1}, \\
&\hat{B} = -\tilde{A}^{-1}\tilde{B}\big(\tilde{D} - \tilde{C}\tilde{A}^{-1}\tilde{B}\big)^{-1}, \\
&\hat{C} = \hat{B}^{T} = -\big(\tilde{D} - \tilde{C}\tilde{A}^{-1}\tilde{B}\big)^{-1}\tilde{C}\tilde{A}^{-1}.\label{invert2}
\end{align}

After the calculation we obtain,

\begin{align}
\hat{A} = &\bigg(\frac{2}{2\delta - i\kappa} \bigg)|0_{c}\rangle\langle 0_{c}|  \nonumber\\ &+ 
\sum_{j=1}^n\Bigg(\frac{2\Big(\Delta - i\gamma\frac{(n+1)}{4} \Big)}{d_{(j-1)}}\Bigg)|j_{c}\rangle\langle j_{c}| 
\end{align}
and

\begin{align}
&\hat{B} = -2g\sum_{j=1}^n \bigg(\frac{\sqrt{j}}{d_{(j-1)}}\bigg)|j_{c}\rangle\langle e (j-1)|, \\
&\hat{C} = \hat{B}^{T}, \\
&\hat{D} = (2\delta - i\kappa)\sum_{j=0}^{n-1}\Big( \frac{1}{d_{j}}\Big)|e (j)\rangle\langle e(j) |,
\end{align}

with,

\begin{equation}
d_{j} = \bigg(\Delta - i\gamma\frac{(n+1)}{4}\bigg)(2\delta - i\kappa) -  2(j+1)g^2.
\end{equation}
It is now  possible to obtain the effective operators, which are found to be

\begin{align}
&\hat{L}^{1}_{eff} = -g\Omega\sqrt{\kappa}\sum_{j=0}^{n-1}\Big( \frac{\sqrt{j+1}\sqrt{n-j}}{d_j}\Big)|j+1\rangle\langle j|, \label{eff1}\\
&\hat{L}^{2}_{eff} = \frac{(2\delta - i\kappa)\Omega\sqrt{\gamma}}{2\sqrt{2}}\sum_{j = 0}^{n-1}\Big( \frac{n-j}{d_j}\Big)|j\rangle\langle j |, \label{eff2}\\
&\hat{L}^{3}_{eff} =\frac{(2\delta - i\kappa)\Omega\sqrt{\gamma}}{2\sqrt{2}}\sum_{j=0}^{n-1}\Big( \frac{\sqrt{j+1}\sqrt{n-j}}{d_j}\Big)|j+1\rangle\langle j|. \label{eff3}
\end{align}
While the effective Hamiltonian is given by

 \begin{equation}\label{effham}
\hat{H}_{eff} = -\frac{\Omega^2}{8}\sum_{j = 0}^{n-1}f(j)(n-j)|j\rangle\langle j|,
\end{equation}
where

\begin{equation}
f(j) = \frac{(2\delta - i\kappa) \overline{d_j} + (2\delta + i\kappa)d_{j}}{|d_{j}|^2}.
\end{equation}

Figure \ref{effsys} summarizes the effective ground state processes and offers excellent insight into the underlying design of the scheme. Each effective process consists of a coherent excitation via laser driving, an intermediate process described by the propagator $\hat{H}^{-1}_{NH}$, and a de-excitation via dissipation or coherent driving.$\hat{L}^{2}_{eff}$ and $\hat{H}_{eff}$  describe effective ``loop" processes, from state $|j\rangle$ to state $|j\rangle$, while $\hat{L}^{1}_{eff}$ and $\hat{L}^{3}_{eff}$ describe state transfer from state $|j\rangle$ to state $|j+1\rangle$, driven by dissipation as the mechanism of de-excitation. From Figure \ref{effsys}, and the form of the effective operators, it is clear that net state transfer is only  possible from left to right, from state $|j\rangle$ to state $|j+1\rangle$, and hence a necessary assumption for this introductory scheme is that the system begins in the state $|0\rangle$.

It is clear that the strength of the effective process from state $|j\rangle$ to state $|j+1\rangle$, described by $\hat{L}^{1}_{eff}$, is determined by the propagator element $h^{1}_{j}$, given by

\begin{equation}
h^{1}_{j} = \langle (j+1)_{c} | \hat{H}^{-1}_{NH} | e (j) \rangle \sim \frac{\sqrt{j+1}}{d_{j}}.
\end{equation}
Similarly, the strength of the effective process from state $|j\rangle$ to state $|j+1\rangle$, described by $\hat{L}^{3}_{eff}$, is determined by the propagator element $h^{3}_{j}$ given by

\begin{equation}
h^{3}_{j} = \langle e (j) | \hat{H}^{-1}_{NH} | e (j) \rangle \sim \frac{1}{d_{j}}.
\end{equation}
Therefore,one can see that in order to prepare the state $|1\rangle$ (the W state of $n$ qubits), it is necessary to choose system parameters such that $d_{0} \ll d_{1}$. This will result in $h^{1}_{0} \gg h^{1}_{1}$ and $h^{3}_{0} \gg h^{3}_{1}$, effectively enhancing the strength of the effective process from state $|0\rangle$ to state $|1\rangle$, while suppressing loss from state $|1\rangle$ into state $|2\rangle$.

In this case the effective master equation can  be solved explicitly, offering extra insight into the manner in which system parameters need to be chosen, and insight into the limitations of the system. The effective master equation consists of $(n+1)^2$ equations for the matrix elements of $\rho_{g}$. For this scheme, the $n+1$ equations for the diagonal matrix elements decouple from the remaining equations, and we are left to solve $n+1$ coupled first-order equations of the form

\begin{equation}\label{eqform}
\dot{\rho}_{j} = T_{(j-1)}\rho_{(j-1)} - T_{j}\rho_{j}, 
\end{equation}
where $\rho_{j} = \langle j |\rho_{g} |j\rangle$, and

\begin{equation}
T_{j} = l^{(1)}_{j} + l^{(3)}_{j},
\end{equation}
with $l^{(1)}_{j}$ and $l^{(3)}_{j}$ defined as

\begin{align}
& l^{(1)}_{j} = |\langle j +1 | \hat{L}^{1}_{eff} | j \rangle |^2, \\
& l^{(3)}_{j} = |\langle j +1 | \hat{L}^{3}_{eff} | j \rangle |^2.  
\end{align}

Our assumption regarding the initial state of the system, along with probability requirements, implies the initial condition 

\begin{equation}
\rho_{j}(t = 0) = \delta_{j,0}.
\end{equation}
It is important to note that $T_{(-1)} = 0$, and that because of the left to right nature of the system, we are only concerned with solving for $\rho_{0}(t)$ and $\rho_{1}(t)$. These solutions are found to be

\begin{align}
&\rho_{0}(t) = e^{-T_{0}t}, \label{soln0}\\
&\rho_{1}(t) = \frac{T_{0}}{T_{0} - T_{1}}\Big[e^{-T_{1}t} - e^{-T_{0}t}   \Big].  \label{soln}
\end{align}
Instantly a heuristic analysis shows that if $T_{0} \gg T_{1}$, then

\begin{align}
&\lim_{t \rightarrow \infty} \rho_{0}(t) = 0, \\
&\lim_{t \rightarrow \infty} \rho_{1}(t) = 1.
\end{align} 

We now focus our attention towards determining the extent to which this can be achieved, and the manner in which system parameters need to be chosen to do this. Using the effective operators we find that

\begin{equation}
T_{j} = g(j)\Big[\Omega^2 \Big(g^2\kappa + \frac{4\delta^2 + \kappa^2}{8}  \Big)   \Big],
\end{equation}
with

\begin{equation}\label{restriction1}
g(j) =  \frac{m(j)}{|d_j|^2} = \frac{(j+1)(n-j)}{|d_j|^2}.
\end{equation}
Hence, in order to achieve $T_{0} \gg T_{1}$ one must have $g_{0} \gg g_{1}$. For small $n$ we have that $m(0) \approx m(1)$ and hence choosing parameters such that $|d_0|^2 \ll |d_1|^2$ will result in $g_{0} \gg g_{1}$ as desired. However, for large $n$ we have that $m(0) \ll m(1)$ and hence, despite achieving $|d_0|^2 \ll |d_1|^2$, we will not be able to achieve $g_{0} \gg g_{1}$ as required. This sets a limit on the size of the W state which can be produced reliably using this scheme - a limit which will be explored shortly.

In order to choose parameters such that $g_{0} \gg g_{1}$, we introduce the following notation,

\begin{equation}
g = y, \qquad \delta = \tilde{\delta}y, \qquad \Delta = \tilde{\Delta}y,
\end{equation}

\begin{equation}
\Omega = \tilde{\Omega} x, \qquad \kappa = \tilde{\kappa}x, \qquad \gamma = \tilde{\gamma}x,
\end{equation}
where $y = \alpha x$, $\alpha \approx 10$ and $(\tilde{\delta},\tilde{\Delta},\tilde{\Omega},\tilde{\kappa}, \tilde{\gamma}) = \mathcal{O}(1)$ enforce the correct scale of each parameter. Utilizing this notation one finds that

\begin{align}
d_{j} =& x^2 \Bigg[ \alpha^2 \big[2(\tilde{\delta}\tilde{\Delta} - (j+1)) \big]   \nonumber\\ &- i\alpha \bigg[\tilde{\Delta}\tilde{\kappa} -\tilde{\gamma}\tilde{\delta} \bigg(\frac{n+1}{2}\bigg) \bigg] 
 - \tilde{\gamma}\tilde{\kappa}\bigg(\frac{n+1}{4}\bigg) \Bigg]
\end{align}

and hence one can approximate $d_{j}$ by

\begin{equation}\label{leadingorder}
d_{j} \approx x^2 \alpha^2 \big[2(\tilde{\delta}\tilde{\Delta} - (j+1)) \big].
\end{equation}
From the above one can see that a parameter choice $\tilde{\delta}\tilde{\Delta} = 1$, yields

\begin{align}
&d_{0} \approx 0, \\
&d_{(j\neq 0)}  \approx - x^2 \alpha^2 j,  
\end{align}
such that one indeed has $|d_0|^2 \ll |d_1|^2$. Again it is important to note that the approximation in Eq. (\ref{leadingorder}) is only valid for small $n$, as for larger $n$ it becomes true that

\begin{equation}\label{restriction2}
\alpha\frac{(n+1)}{2} \approx \alpha^2
\end{equation}
and hence the assumption breaks down. 

It is now possible to derive relevant benchmarks, allowing for a detailed examination of the protocol. From Eq. \eqref{soln} one finds 

\begin{equation}
T_{p} = \frac{\mathrm{ln} \Big(\frac{T_{1}}{T_{0}}\Big)}{(T_1 - T_0)},
\end{equation}
where $T_{p}$ is the time taken to reach $\rho_{(1,\mathrm{max})}$, the maximum population of $\rho_{1}$ obtained, given by

\begin{figure} 
\includegraphics[width=0.9\linewidth]{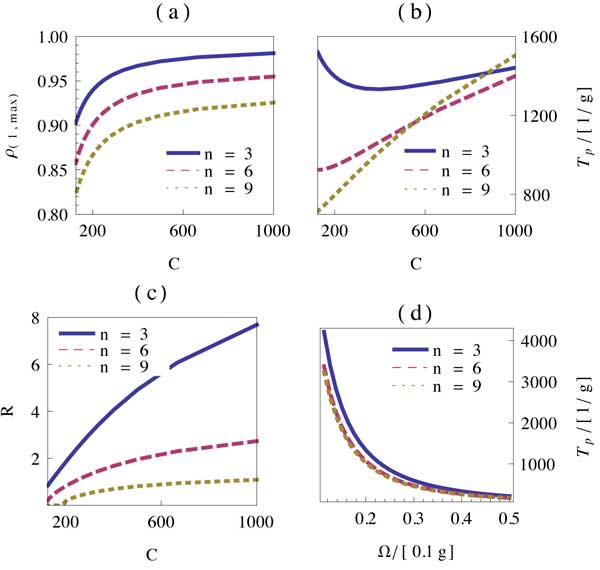} 
\caption{(Color online) (a-c) Plots of protocol benchmarks against cooperativity $C$, for different values of $n$. For these plots $(\tilde{\Omega,\tilde{\gamma}}) = (1/5,1/2)$ such that the protocol is within the weak driving regime and the cooperativity is varied through $\tilde{\kappa}$. (d) Plot of preparation time, $T_{p}$, as a function of coherent driving strength $\Omega$ for different values of $n$. For this plot $(\tilde{\kappa,\tilde{\gamma}}) = (1/2,1/2)$ such that $C=400$. (a-d) For all plots $(\tilde{\delta},\tilde{\Delta}) = (8/11,11/8)$, the numerically optimized choice within the restriction $\tilde{\Delta}\tilde{\delta} = 1$, while $z = 0.85$. Furthermore, typical values of $g$ are on the order of 10 MHz \cite{kimbleQED}, such that preparation times are on the order of $\mu s$, and $\Omega$ values are on the order of MHz.}\label{SCblock} 
\end{figure}

\begin{equation}\label{bench1}
\rho_{(1,\mathrm{max})} = \rho_{1}(T_{p}).
\end{equation} 

It is also of interest to examine the stability time of the target state given by the expression

\begin{equation}
S = T_{z} - T_{p},
\end{equation}
with $T_{z}$ the time, $T_{z} > T_{p}$, such that

\begin{equation}
\rho_{1}(T_{z}) = z,
\end{equation}
where $z$ is some specified population, set in order to measure the threshold decay of the target state. Finally, it is of interest to examine the ratio of the stability time of the target state to the preparation time, given by the expression

\begin{equation}\label{bench2}
R = \frac{S}{T_{p}}.   
\end{equation}

It is of particular interest to explore the behaviour of the above benchmarks with respect to cooperativity (a dimensionless and invariant measure of the quality of a cavity QED system) and system size, where cooperativity for a single cavity is given by the expression,

\begin{equation}
C = \frac{g^2}{\kappa\gamma}.
\end{equation}

 Figure \ref{SCblock} (a-c) details the behavior of primary benchmarks $(\rho_{(1,\mathrm{max})}, R, T_{p})$ with respect to the relevant parameters of cooperativity and system size. For all displayed results the values of $\tilde{\delta}$ and $\tilde{\Delta}$ utilized have been numerically chosen to maximize the ratio $T_{0}/T_{1}$, based on the criterion $\tilde{\Delta}\tilde{\delta} = 1$, but taking into account the lower order terms in $\alpha$ of $d_{j}$. It is also important to note that $\Delta$ and $\delta$ satisfy the conditions necessary for adiabatic elimination, while $\Omega$ is well within the regime of weak driving in which the accuracy of the effective operator formalism has been thoroughly analysed and firmly established in \cite{effop}.  
 
It is clear that for currently available cooperativities \cite{prep1} in the range $C \approx 200$, and small system sizes corresponding to $n \approx 3$, the protocol behaves comparably to the previously suggested protocols \cite{prepW} for $n=3$, with only one laser necessary in this case. Importantly, the behaviour of the system with respect to all benchmarks, at a fixed system size, scales excellently with respect to cooperativity. For increased cooperativities, the realization of which is an active field of current research, it is possible to obtain effective steady states (states with extremely slow decay) with fidelities of near unity and rapid preparation times. It is important to note, from Figure \ref{SCblock} (d), that within the regime of weak driving it is possible to obtain a broad range of preparation times. It can be shown that $R$ and $\rho_{(1,\mathrm{max})}$ exhibit no dependence on $\Omega$ such that variation of coherent driving strength allows for the preparation of extremely stable states, even at low cooperativities, at the cost of increased preparation times.

The strength of this protocol, and the motivation for this work, is the ease at which it is possible to scale the  protocol to larger system sizes, without the need for any additional lasers. From our analysis, especially expressions \eqref{restriction1} and \eqref{restriction2}, we do not expect the protocol to succeed for arbitrarily large systems, however Figure \ref{SCblock} shows that reasonable scaling, up to system sizes of $n\approx 10$ is possible, an order of magnitude improvement over previous schemes. It is clear that within a fixed cooperativity the performance of the protocol, with respect to all benchmarks, decreases as a function of system size. However, scaling of the performance with cooperativity is not affected, such that these decreases in performance can be combated via the utilization of larger cooperativity cavities, envisaged experimentally possible in the near future.

\begin{figure} 
\includegraphics[width=0.7\linewidth]{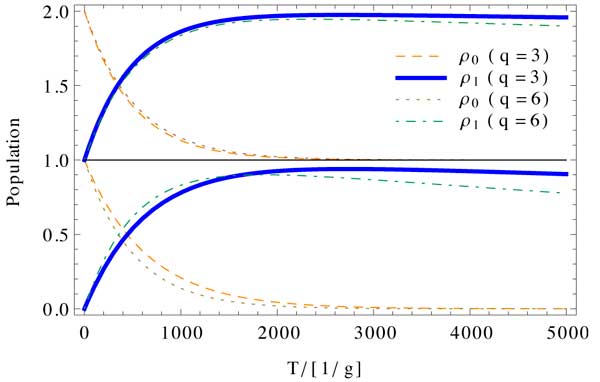} 
\caption{(Color online) Evolution of populations with time, from initial state $|0\rangle$, for different values of cooperativity, and different system sizes. The lower plot corresponds to $C=200$ with $(\tilde{\gamma},\tilde{\kappa}) = (1/2,1)$ while the upper plot corresponds to $C=600$, with $(\tilde{\gamma},\tilde{\kappa}) = (1/2,1/3)$. Both plots are at $(\tilde{\Omega},\tilde{\delta},\tilde{\Delta}) = (1/5,8/11,11/8)$. Note that typical values of $g$ are on the order of 10 MHz \cite{kimbleQED}, such that preparation times are on the order of $\mu s$. Populations in the upper plot have been increased by one for display. }\label{evolution} 
\end{figure}

The evolution of populations is shown in Figure \ref{evolution} where it is clear that effective steady states are easily produced for small system sizes, even at low cooperativites, while for larger system sizes increased cooperativites are necessary in order to obtain long lived states of high fidelity. However, at slightly higher cooperativites results are obtained for large system sizes, comparable to those previously obtained for $n=3$  \cite{prepW}.

\subsection{Bimodal Cavity scheme}

In the previous section we presented a scheme with excellent properties, under the assumption that the system starts in the thermal ground state $|0\rangle$. This assumption was necessary, as from Figure \ref{effsys} and the form of the effective operators \eqref{eff1} - \eqref{eff3}, it is clear that effective state transfer, driven by dissipation, was only possible from ``left to right", or from state $|j\rangle$ to state $|j+1\rangle$. In order to construct a scheme in which the ideas of the previous scheme are exploited, but it is possible to start in any thermal state of the system, it is necessary to consider a physical system which results in the possibility of effective bi-directional state transfer. 

\begin{figure} 
\begin{center}
\includegraphics[width=0.7\linewidth]{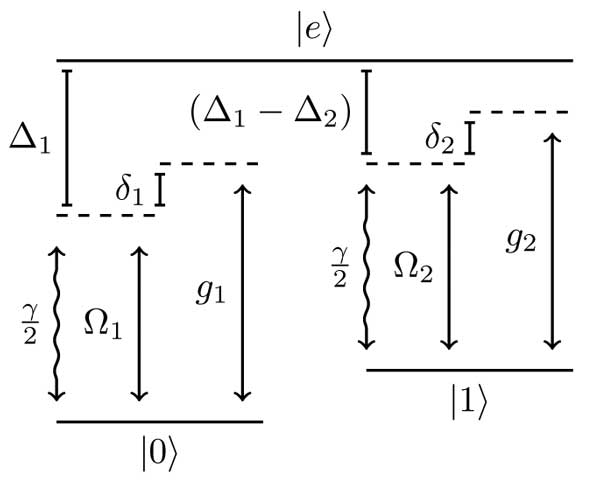} 
 \caption{Cavity QED setup for a single atom in a bimodal cavity. The entire set up consists of $n$ identical atoms within a single cavity.}\label{bimodalatom} 
 \end{center}
\end{figure}

Bimodal cavities, as studied in \cite{bimodal1}-\cite{bimodal3}, offer the perfect physical realization of such a system. Dissipation processes involving one mode of the cavity can be utilized to drive effective ``left to right" processes, while another dissipation process involving the other mode of the cavity can be utilized to drive effective ``right to left" processes. We consider a bimodal cavity QED set up of $n$ $\Lambda$ atoms in a bimodal cavity, as illustrated in Figure \ref{bimodalatom}

In this case the total Hilbert space for $n$ $\Lambda$ atoms, in a cavity with two modes, each mode restricted to a single excitation, is given by

\begin{equation}
\mathcal{H} = \bigg( \bigotimes_{i = 1}^{n} \mathbb{C}^3 \bigg)\otimes\mathbb{C}^2\otimes\mathbb{C}^2,
\end{equation}
where the full Hilbert space of the $i$'th atom is $\mathbb{C}^3$ and spanned by the basis $\{|0\rangle_{i},|1\rangle_{i},|e\rangle_{i}\}$, the Hilbert space of the first cavity field mode, restricted to a single excitation and with creation and annihilation operators $(\hat{a}^{\dagger},\hat{a})$, is $\mathbb{C}^2$ and spanned by the basis $\{|0\rangle_{c(1)},|1\rangle_{c(1)}\}$ and the Hilbert space of the second cavity field mode, restricted to a single excitation and with creation and annihilation operators $(\hat{b}^{\dagger},\hat{b})$, is $\mathbb{C}^2$ and spanned by the basis $\{|0\rangle_{c(2)},|1\rangle_{c(2)}\}$.

The Hamiltonian for the system at hand, in the conventional basis and the appropriate rotating frame, is again time independent with the form of \eqref{hamform}, and individual elements given by

\begin{align}
&\hat{H}_{e} = \Delta_{1}\sum_{j = 1}^{n}|e\rangle_{j}\langle e | + \delta_{1}(\hat{a}^{\dagger}\hat{a}) + \delta_{2}(\hat{b}^{\dagger}\hat{b}) + \hat{H}_{ac}, \label{h1}\\
& \hat{H}_{ac} = g_{1}\Big(\hat{a}^{\dagger}\hat{V}_{+} + H.C  \Big) + g_{2}\Big(\hat{b}^{\dagger}\hat{U}_{+} + H.C  \Big), \\
& \hat{W}_{+} = \bigg(\frac{\Omega_{1}}{2}\bigg)  \hat{V}_{-} + \bigg(\frac{\Omega_{2}}{2}\bigg) \hat{U}_{-}, \\
& \hat{W}_{-} = \hat{W}_{+}^{\dagger}, \\
& \hat{H}_{g} = \Delta_{2}\sum_{j = 1}^{n}|1\rangle_{j}\langle 1 |. \label{h2}
\end{align}
Coherent laser driving with a resonant Rabi frequency of $\Omega_{1}$ and a detuning of $\Delta_{1}$ is applied uniformly over all atoms and couples the levels $|0\rangle$ and $|e\rangle$, while coherent laser driving with a resonant Rabi frequency of $\Omega_{2}$ and a detuning of $\Delta_{1} - \Delta_{2}$ couples the levels $|1\rangle$ and $|e\rangle$, also uniformly over all atoms. The levels $|0\rangle$ and $|e\rangle$ are also coupled via the cavity field $(\hat{a}^{\dagger},\hat{a})$, with a strength of $g_{1}$ and uniform phase over all atoms, where a cavity excitation of the first field mode, created by $\hat{a}^{\dagger}$, has an energy of $\delta_{1}$. Finally, the levels $|1\rangle$ and $|e\rangle$ experience coupling via the cavity field $(\hat{b}^{\dagger},\hat{b})$, with a strength of $g_{2}$ and uniform phase over all atoms, where a cavity excitation of the second field mode, created by $\hat{b}^{\dagger}$, has an energy of $\delta_{2}$.

Again the system interacts with the environment via spontaneous emission and cavity loss. Under the assumption that these dissipation processes are Markovian, the system is described by a master equation with Lindblad operators

\begin{align}
&\hat{L}_{(\gamma,0)} = \sqrt{\frac{\gamma}{2}}\hat{V}_{+} = \sqrt{\frac{\gamma}{2}}\sum_{i = 1}^n|0\rangle_{i}\langle e |, \label{l1}\\
&\hat{L}_{(\gamma,1)} = \sqrt{\frac{\gamma}{2}}\hat{U}_{+}=\sqrt{\frac{\gamma}{2}} \sum_{i = 1}^n|1\rangle_{i}\langle e |, \\
& \hat{L}_{\kappa_{1}} = \sqrt{\kappa_{1}}\hat{a}, \\
& \hat{L}_{\kappa_{2}} = \sqrt{\kappa_{2}}\hat{b}, \label{l2}
\end{align}
where $\kappa_{1}$ is the photon decay rate for the cavity field $(\hat{a}^{\dagger},\hat{a})$ and $\kappa_{2}$ is the photon decay rate for the cavity field $(\hat{b}^{\dagger},\hat{b})$. Again the spontaneous decay rates into states $|0\rangle$ and $|1\rangle$ have been set equal for simplicity. Following from the previous example, we would like to utilize the fully-symmetric single excitation basis to describe our Hamiltonian and Lindblad operators. This basis, spanning the fully-symmetric single excitation subspace of the total Hilbert space, is given by

\begin{equation}
B = \{G,A,C_{1},C_{2}\}.
\end{equation} 
$G$ is the fully symmetric ground state basis, as per \eqref{G}, $A$ is the fully symmetric single atomic-excitation basis, as per Eq. \eqref{A} and we utilize the natural notation

\begin{align} \label{notnew}
&|j\rangle = |j\rangle\otimes|0_{c(1)}\rangle \otimes|0_{c(2)}\rangle, \\
&|e(j)\rangle = |e(j)\rangle\otimes|0_{c(1)}\rangle \otimes|0_{c(2)}\rangle.
\end{align}
$C_{i}$ is the fully symmetric single cavity-excitation basis for cavity mode $i$, where

\begin{equation}
C_{i} = \left\{ |0_{c(i)}\rangle, \ldots ,  \frac{1}{\sqrt{ \binom{n}{j} } }|j_{c(i)}\rangle , \ldots, |n_{c(i)}\rangle  \right\}
\end{equation}
and the state $|j_{c(i)}\rangle$ is given by

\begin{equation}
|j_{c(1)}\rangle = \hat{a}^{\dagger}|j\rangle, \qquad |j_{c(2)}\rangle = \hat{b}^{\dagger}|j\rangle.
\end{equation}

In light of the previous discussion it is clear that the fully symmetric single-excitation subspace, containing the W state, is closed under the action of the Hamiltonian \eqref{h1} - \eqref{h2} and Lindblad operators \eqref{l1} - \eqref{l2}. Therefore, as before, we restrict ourselves to this subspace and proceed by transforming the Hamiltonian and Lindblad operators into the basis $B$. This results in a Hamiltonian where $\hat{H}_{e}$ is given  by

\begin{align}
\hat{H}_{e} =& \sum_{j = 0}^{n}\bigg[\Big(\Delta_{1} + j\Delta_{2}\Big)|e(j)\rangle\langle e(j)| \nonumber\\
 &+ j\Delta_{2}\Big(|j_{c(1)}\rangle\langle j_{c(1)}| + |j_{c(2)}\rangle\langle j_{c(2)}| \Big) \bigg] + \hat{H_{ac}},
\end{align}

with the atom-cavity interaction Hamiltonian

\begin{align}
\hat{H}_{ac} =& g_{1}\bigg( \sum_{j=0}^{n}(\sqrt{n-j})| j_{c(1)}\rangle\langle e(j)| + h.c.  \bigg) \nonumber\\ + 
& g_{2}\bigg( \sum_{j=0}^{n-1}(\sqrt{j+1})| (j+1)_{c(2)}\rangle\langle e(j)| + h.c.  \bigg).
\end{align}
The perturbative excitation term is given by

\begin{align}
\hat{W}_{+} =&\bigg(\frac{\Omega_{1}}{2}\bigg)\bigg(\sum_{j=0}^{n}(\sqrt{n-j})|e(j)\rangle\langle j | \bigg) \nonumber\\ &+
\bigg(\frac{\Omega_{2}}{2}\bigg)\bigg(\sum_{j=0}^{n}(\sqrt{j+1})|e(j)\rangle\langle (j+1) | \bigg)
\end{align}
and finally, the ground state Hamiltonian is given by

\begin{equation}
\hat{H}_{g} = \sum_{j=0}^{n}(\Delta_{2}j)|j\rangle\langle j|.   
\end{equation}

In order to calculate $\hat{H}_{NH}$ one requires the Lindblad operators, which after a basis transformation are found to be,

\begin{align}
&\hat{L}_{1} = \hat{L}_{\kappa_{1}} = \sqrt{\kappa_{1}}\sum_{j= 0}^{n}|j\rangle\langle j_{c(1)}|, \\
&\hat{L}_{2} = \hat{L}_{\kappa_{2}} = \sqrt{\kappa_{2}}\sum_{j= 0}^{n}|j\rangle\langle j_{c(2)}|, \\
&\hat{L}_{3} = \hat{L}_{\kappa_{(\gamma,1)}} = \sqrt{\frac{\gamma}{2}}\sum_{j= 0}^{n-1}(\sqrt{j+1})|j +1 \rangle\langle e(j)|, \\
&\hat{L}_{4} = \hat{L}_{\kappa_{(\gamma,0)}} = \sqrt{\frac{\gamma}{2}}\sum_{j= 0}^{n}(\sqrt{n-j})|j  \rangle\langle e(j)|.
\end{align}

\begin{figure}[H] 
\begin{center}
\includegraphics[width=0.7\linewidth]{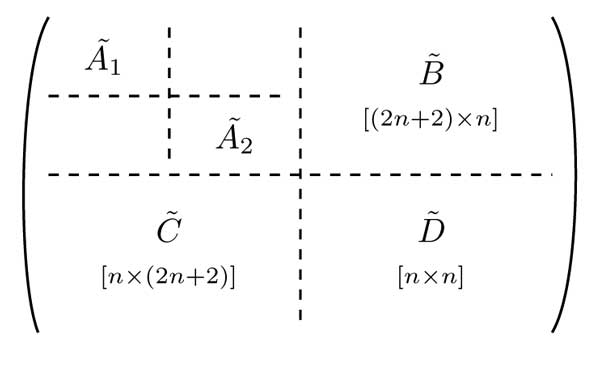} 
 \caption{Partitioned matrix form of $\hat{H}_{NH}$ for the Bimodal scheme. }\label{BM2}
 \end{center}
\end{figure}

All of the above now allows us to calculate the non-hermitian Hamiltonian as per \eqref{HNHdef}. As in the previous scheme, in order to invert $\hat{H}_{NH}$ it is useful to represent this operator as a partitioned matrix with form as per Figure \ref{BM2}.

In this case,

\begin{equation}
\hat{H}_{NH} = \tilde{A}_{1} + \tilde{A}_{2} + \tilde{B} + \tilde{C} + \tilde{D}.
\end{equation}

We define $\tilde{A} = \tilde{A}_{1} + \tilde{A}_{2}$ with

\begin{align}
& \tilde{A}_{1} = \sum_{j = 0}^{n}\bigg[ \Big(\delta_{1} + j\Delta_{2} \Big) - i\frac{\kappa_{1}}{2}\bigg]| j_{c(1)}\rangle\langle j_{c(1)} |, \\
& \tilde{A}_{2} = \sum_{j = 0}^{n}\bigg[ \Big(\delta_{2} + j\Delta_{2} \Big) - i\frac{\kappa_{2}}{2}\bigg]| j_{c(2)}\rangle\langle j_{c(2)} |, \\
\end{align}
while the remaining blocks are

\begin{align}
\tilde{B} =& \sum_{j = 0}^{n-1} \bigg[ g_{1}(\sqrt{n-j})| j_{c(1)}\rangle\langle e(j)|  \nonumber\\& \quad+  g_{2}(\sqrt{j+1})| (j+1)_{c(2)}\rangle\langle e(j)|    \bigg],
\end{align}

\begin{equation}
\tilde{D} = \sum_{j = 0}^{n-1}\bigg[ \Big(\Delta_{1} + j\Delta_{2} \Big) - i\gamma\frac{(n+1)}{4}\bigg]| e(j)\rangle\langle e(j) |,
\end{equation}
and $\tilde{C} = \tilde{B}^{T}$. Using the Banachiewicz inversion theorem, via Eqs. \eqref{invert1} - \eqref{invert2}, we obtain

\begin{equation}
\hat{H}_{NH}^{-1} = \hat{A} + \hat{B} + \hat{C} + \hat{D}.
\end{equation}
Here $\hat{A} = \hat{A}_{1} + \hat{A}_{2} + \hat{A}_{3}$, where

\begin{align}
&\hat{A}_{1} = \sum_{j = 0}^{n} f_{1}(j) |j_{c(1)}\rangle\langle j_{c(1)} |, \\
&\hat{A}_{2} = \bigg(\frac{2}{\alpha^{(2)}_{0}}\bigg) |0_{c(2)}\rangle\langle 0_{c(2)} | \nonumber\\
&\qquad\qquad +  \sum_{j = 0}^{n} f_{2}(j) |(j+1)_{c(2)}\rangle\langle (j+1)_{c(2)} |, \\
&\hat{A}_{3} = \sum_{j = 0}^{n} f_{3}(j) (|j_{c(1)}\rangle\langle (j+1)_{c(2)}| + h.c.)
\end{align}
and we have defined the functions

\begin{align}
&f_{1}(j) = \frac{2d_{j} - 16g_{1}^2(n-j)\alpha^{(2)}_{j+1}\alpha^{(1)}_{j}}{\alpha^{(1)}_{j}d_{j}}, \\
&f_{2}(j) = \frac{2d_{j} - 16g_{2}^2(j+1)\alpha^{(2)}_{j+1}\alpha^{(1)}_{j}}{\alpha^{(2)}_{j+1}d_{j}}, \\
&f_{3}(j) = \frac{(16g_{1}g_{2}\sqrt{n-j}\sqrt{j+1})}{d_{j}},
\end{align}
with $\alpha^{(k)}_{j}$ defined via

\begin{equation}
\alpha^{(k)}_{j} = 2(\delta_{k} + j\Delta_2) - i\kappa_{k},
\end{equation}
and $d_{j}$ defined via

\begin{align}
d_{j} =& \beta_{j}\alpha^{(1)}_{j}\alpha^{(2)}_{j+1}  \nonumber\\ &- 8\Big[g^2_{1}(n-j)\alpha^{(2)}_{j+1} + g^2_{2}(j+1)\alpha^{(1)}_{j}\Big],
\end{align}
with $\beta_{j}$ given by

\begin{equation}
\beta_{j} = 4(|\Delta_{1} + j\Delta_{2}) - i\gamma(n+1).
\end{equation}
The block containing propagators between atomic and cavity single-excitation states is given by $\hat{B} = \hat{B}_{1} + \hat{B}_{2}$, with

\begin{align}
&\hat{B}_{1} = \sum_{j = 0}^{n-1}\bigg(\frac{-8 g_{1}(\sqrt{n-j})\alpha^{(2)}_{j+1}}{d_{j}}  \bigg) |j_{c(1)}\rangle\langle e(j)|,   \\
&\hat{B}_{2} = \sum_{j = 0}^{n-1}\bigg(\frac{-8 g_{2}(\sqrt{j+1})\alpha^{(1)}_{j}}{d_{j}}  \bigg) |(j+1)_{c(2)}\rangle\langle e(j)|, 
\end{align}
and $\hat{C} = \hat{B}^{T}$. Finally, the block containing loop propagators between atomic single-excitation states is given by

\begin{equation}
\hat{D} = \sum_{j=0}^{n-1} \bigg(\frac{4\alpha^{(1)}_{j}\alpha^{(2)}_{j+1}}{d_{j}}\bigg)|e(j)\rangle\langle e(j)|.
\end{equation}

Armed with $H^{-1}_{NH}$ it is possible to calculate the effective operators utilizing Eqs. \eqref{heffdef} and \eqref{leffdef}. For all effective operators we find that $\hat{L}^{i}_{eff} = \hat{L}^{i(a)}_{eff}  +\hat{L}^{i(b)}_{eff}$. The results are as follows;

\begin{align}
&\hat{L}_{eff}^{1(a)} = c_{1}(1)\sum_{j = 0}^{n-1}\bigg(\frac{(n-j)\alpha^{(2)}_{j+1}}{d_{j}}\bigg)|j\rangle\langle j|,   \label{efflist1}\\ 
&\hat{L}_{eff}^{1(b)} = c_{1}(2)\sum_{j = 0}^{n-1}\bigg(\frac{(n-j)(j+1)\alpha^{(2)}_{j+1}}{d_{j}}\bigg)|j\rangle\langle j+1|, \\
&\hat{L}_{eff}^{2(a)} = c_{2}(1)\sum_{j = 0}^{n-1}\bigg(\frac{(n-j)(j+1)\alpha^{(1)}_{j}}{d_{j}}\bigg)|j+1\rangle\langle j|, \\
&\hat{L}_{eff}^{2(b)} = c_{2}(2)\sum_{j = 0}^{n-1}\bigg(\frac{(j+1)\alpha^{(1)}_{j}}{d_{j}}\bigg)|j+1\rangle\langle j+1|, \\
&\hat{L}_{eff}^{3(a)} = c_{3}(1)\sum_{j = 0}^{n-1}\bigg(\frac{\sqrt{n-j}\sqrt{j+1}\alpha^{(1)}_{j}\alpha^{(2)}_{j+1}}{d_{j}}\bigg)|j+1\rangle\langle j|, \\
&\hat{L}_{eff}^{3(b)} = c_{3}(2)\sum_{j = 0}^{n-1}\bigg(\frac{(j+1)\alpha^{(1)}_{j}\alpha^{(2)}_{j+1}}{d_{j}}\bigg)|j+1\rangle\langle j+1|,\\
&\hat{L}_{eff}^{4(a)} = c_{3}(1)\sum_{j = 0}^{n-1}\bigg(\frac{(n-j)\alpha^{(1)}_{j}\alpha^{(2)}_{j+1}}{d_{j}}\bigg)|j\rangle\langle j|,\\
&\hat{L}_{eff}^{4(b)} = c_{3}(2)\sum_{j = 0}^{n-1}\bigg(\frac{\sqrt{n-j}\sqrt{j+1}\alpha^{(1)}_{j}\alpha^{(2)}_{j+1}}{d_{j}}\bigg)|j\rangle\langle j+1|, \label{efflistlast}
\end{align}
where the constants are given by

\begin{figure*} 
\includegraphics[width=0.9\linewidth]{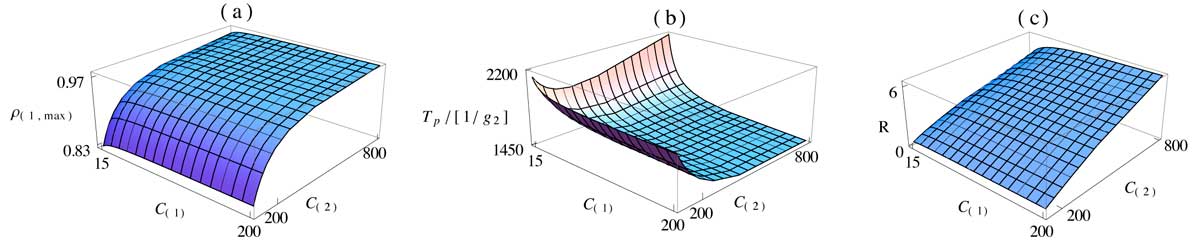} 
\caption{(Color online) (a-c) Plots of protocol benchmarks against cooperativity values $C_{(1)}$ and $C_{(2)}$, for fixed system size with $n=3$ and for the case $\Omega_{2} = 0$. For these plots $(\tilde{\Omega,\tilde{\gamma}}) = (1/5,1/2)$ such that the protocol is within the weak driving regime and cooperativities are varied through $\tilde{\kappa_{1}}$ and $\tilde{\kappa_{2}}$. For all plots $(\delta_{1},\delta_{2},\Delta_{1},g_{1})$ are at numerically optimized values, as per Eq. \eqref{optimum}, within the restrictions set by Eqs. \eqref{paramch1} and \eqref{paramch2}. The threshold value has been set strictly, with $z = 0.85$ and the initial state of the system is assumed as $|0\rangle$. Furthermore, typical values of $g$ are on the order of 10 MHz \cite{kimbleQED}, such that preparation times are on the order of $\mu s$.}\label{BC1} 
\end{figure*}

\begin{align}
&c_{1}(k) = (-4\sqrt{\kappa_{1}}\Omega_{k}g_{1}), \label{c1}\\
&c_{2}(k) = (-4\sqrt{\kappa_{2}}\Omega_{k}g_{2}), \\
&c_{3}(k) = \sqrt{2}\sqrt{\gamma}\Omega_{k}.      \label{c3}
\end{align}

As can be seen from Eqs. \eqref{efflist1} - \eqref{efflistlast}, there now exist effective loop processes (from state $|j\rangle$ to state $|j\rangle$), effective left to right processes (from state $|j\rangle$ to state $|j+1\rangle$) and effective right to left processes (from state $|j+1\rangle$ to state $|j\rangle$). Closer inspection of the constants in Eqs. \eqref{c1} -\eqref{c3} shows that right to left effective processes, due to $\hat{L}_{eff}^{1(b)}$ and $\hat{L}_{eff}^{4(b)}$, involve a coherent excitation via $\Omega_{2}$, followed by an intermediate process governed by $\hat{H}^{-1}_{NH}$, and a de-excitation via a dissipative process. However, left to right effective processes, due to $\hat{L}_{eff}^{2(a)}$ and $\hat{L}_{eff}^{3(a)}$, involve coherent excitation via  $\Omega_{1}$ as a first step, before intermediate propagation and de-excitation due to dissipation. Therefore, it is clear that 

\begin{equation}
\Omega_{2} = 0 \Rightarrow c_{i}(2) = 0,
\end{equation}
such that only effective left to right processes remain, driven by spontaneous emission and dissipation involving the first cavity-field mode. This is completely analogous to the single-mode cavity scheme, studied in detail in the previous section. However, it is also clear that

\begin{equation}
\Omega_{1} = 0 \Rightarrow c_{i}(1) = 0,
\end{equation}
such that only effective right to left processes remain, driven by spontaneous emission and dissipation involving the second cavity-field mode. In this case the natural steady state of the system, irrespective of initial state, is the state $|0\rangle$. 
 
It is now clear that if it is possible to create long lived W states, with $\Omega_{2} = 0$ and assuming the initial state of the system as $|0\rangle$, then it is possible to create a W state irrespective of initial state, by utilizing a two step process with the first step creating the the state $|0\rangle$ by setting $\Omega_{1} = 0$. We proceed to demonstrate a method for the production of long lived W states, assuming the initial state of the system as $|0\rangle$ and with $\Omega_{2} = 0$, before exploring the initial preparation of the state $|0\rangle$ with $\Omega_{1} = 0$.

In the case of $\Omega_{2} = 0$, our system is completely analogous to the single-mode scheme explored in the previous section. Again, the $n+1$ coupled differential equations for the diagonal elements of the effective master equation decouple from the equations for the off diagonal elements, and we are left to solve $n+1$ coupled differential equations with the form of Eq. \eqref{eqform}. In this case $\rho_{j} = \langle j |\rho_{g} |j\rangle$, and

\begin{equation}
T_{j} = l^{2(a)}_{j} + l^{3(a)}_{j},
\end{equation}
with $l^{2(a)}_{j}$ and $l^{3(a)}_{j}$ defined as

\begin{align}
& l^{2(a)}_{j} = |\langle j +1 | \hat{L}^{2(a)}_{eff} | j \rangle |^2, \\
& l^{3(a)}_{j} = |\langle j +1 | \hat{L}^{3(a)}_{eff} | j \rangle |^2.  
\end{align}
Again, because of the strictly left to right nature of the system we are only concerned with solutions for $\rho_{0}$ and $\rho_{1}$, which are given by Eqs. \eqref{soln0} and \eqref{soln}. It is clear that again the extent to which the ratio $T_{0}/T_{1}$ can be maximised determines the effectiveness of the scheme. For this physical set up we find that

\begin{equation}
T_{j} = h(j)\Big[\Omega_{1}^2|\alpha^{(1)}|^2 \big(16\kappa_{2}g_{2}^2 + 2\gamma|\alpha^{(2)}|^2 \big)\Big],
\end{equation}
where we have set $\Delta_{2} = 0$ such that $\alpha^{(k)}_{j} = \alpha^{(k)}$ no longer depends on $j$, and

\begin{equation}
h(j) = \frac{m(j)}{|d_{j}|^2} = \frac{(n-j)(j+1)}{|d_{j}|^2}.
\end{equation}
Identically to the previous analysis, for small $n$ we find that the denominator of the propagators is the crucial element, and that

\begin{equation}
|d_{0}|^2 \ll |d_{1}|^2 \Rightarrow  T_{0} \gg T_{1}.
\end{equation}
For this physical set up,

\begin{align}
d_{j} =& x^3 \Big[\alpha^{3} d^{(1)}(j) + \alpha^{2} d^{(2)}(j) \nonumber\\&\qquad\qquad+ \alpha  d^{(3)}(j) + d^{(4)}(j)\Big],
\end{align}
where we have defined $y = \alpha x$, with $\alpha \approx 10$ and

\begin{align}
&g_{2} = y, \qquad\quad g_{1} = \tilde{g_{1}}y,    \\
&\delta_{1} = \tilde{\delta_{1}}y, \qquad \delta_{2} = \tilde{\delta_{2}}y,  \\
&\Delta_{1} = \tilde{\Delta_{1}}y, \quad \Delta_{2} = \tilde{\Delta_{2}}y,
\end{align}
defines the large parameters, with

\begin{align}
&\Omega_{1} = \tilde{\Omega_{1}} x, \qquad \Omega_{2} = \tilde{\Omega_{2}} x,  \\
&\kappa_{1} = \tilde{\kappa_{1}}x, \qquad \kappa_{2} = \tilde{\kappa_{2}}x, \\
&\gamma = \tilde{\gamma}x,
\end{align}
defining the small parameters. In terms of the above, the largest term of $d_{j}$ (with $\tilde{\Delta_{2}} = 0$ already set for simplicity) is given by

\begin{equation}
d^{(1)}(j) = 16\Big[ \tilde{\delta_{1}}\tilde{\delta_{2}}\tilde{\Delta_{1}}   - \tilde{g_{1}}^2\tilde{\delta_{2}}(n-j) - \tilde{\delta_{1}}(j+1)\Big],
\end{equation}
such that for parameter choices,

\begin{equation}\label{paramch1}
\tilde{g}_{1} = \frac{1}{b}, \quad \tilde{\delta}_{1} = \frac{\delta_{2}}{a},
\end{equation}
\begin{equation}\label{paramch2}
\tilde{\Delta}_{1} = \frac{1}{c}, \quad \tilde{\delta}_{2} = c\bigg(1 + n\frac{a}{b^2}  \bigg),
\end{equation}
we obtain that

\begin{align}
&d^{(1)}(0) = 0, \\
&d^{(1)}(j \neq 0) = 16 j c \bigg(1 + n\frac{a}{b^2}  \bigg)\bigg(\frac{1}{b^2} - \frac{1}{a^2} \bigg).
\end{align}
Utilizing Eqs. \eqref{paramch1} and \eqref{paramch2}, numerical optimization taking into account lower order terms in $\alpha$ of $d_{j}$, finds that parameter choices,

\begin{equation}\label{optimum}
a = \frac{1}{4}, \qquad b = 2, \qquad c = \frac{1}{2},
\end{equation}
maximise the ratio $T_{0}/T_{1}$, while still yielding values for $\Delta_{1},\delta_{1},\delta_{2}$ and $g_{1}$ corresponding to a physically implementable system in which adiabatic elimination of excited levels can be applied.

\begin{figure} 
\includegraphics[width=0.8\linewidth]{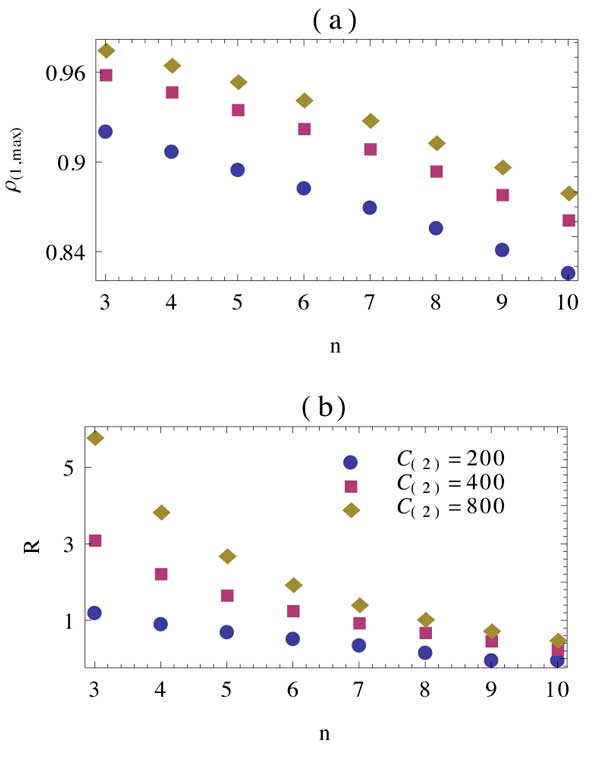} 
\caption{(Color online) (a-b) Plots of protocol benchmarks against system size at different values of cooperativity $C_{(2)}$, for the case $\Omega_{2} = 0$. For these plots $(\tilde{\Omega,\tilde{\gamma}},\tilde{\kappa_{1}}) = (1/5,1/2,1)$ such that the protocol is within the weak driving regime and $C_{2}$ is varied through $\tilde{\kappa_{2}}$ while $C_{1} = 100$ remains fixed. For all plots $(\delta_{1},\delta_{2},\Delta_{1},g_{1})$ are at numerically optimized values, as per Eq. \eqref{optimum}, within the restrictions set by Eqs. \eqref{paramch1} and \eqref{paramch2}. The threshold value has been set strictly, with $z = 0.85$, and the system is assumed to be in the initial state $|0\rangle$.}\label{BC2} 
\end{figure}

It is now possible to examine the behavior of the system, with respect to all relevant benchmarks, as per Eqs. \eqref{bench1} - \eqref{bench2}. For all analysis, the optimum parameter values given in Eq. \eqref{optimum} are utilized. In this bimodal scheme, all benchmarks are a function of $\Omega_{1}$, $C_{(1)}$ , $C_{(2)}$ and $n$, where

\begin{equation}
C_{(i)} = \frac{g_{i}^2}{\kappa_{i}\gamma}
\end{equation} 
is the co-operativity pertaining to the $i$'th cavity mode. From Figure \ref{BC1} it is clear that the behavior of all benchmarks, for a small system with size $n=3$, is extremely similar to the behavior previously observed in the single mode scheme, with respect to $C_{2}$, while the dependence of all benchmarks on $C_{(1)}$ is extremely weak. This is as to be expected by virtue of the fact that dissipation involving the second cavity field mode is the driving mechanism for ``left to right" processes. Again, for low cooperativities one can obtain results, with respect to all benchmarks, comparable to those from previously suggested schemes \cite{prepW}, while the scaling with cooperativity is favourable such that the performance of the protocol can be greatly increased via the use of larger cooperativity cavities. In terms of dependence on $\Omega_{1}$, it can be shown that both $R$ and $\rho_{(1,max)}$ again exhibit no dependency, while the dependency of $T_{p}$ is extremely similar to that from Figure \ref{SCblock} (d), such that a broad range of preparation times can be achieved within the necessary regime of weak driving.

For the Bimodal scheme the system again decreases in performance with respect to all benchmarks, at fixed cooperativities, with increasing system size. However, as can be seen from Figure \ref{BC2}, it is again possible to combat this decrease in performance through an increase in the relevant cooperativity, $C_{(2)}$. The evolutions for this scheme are extremely similar to those shown in Figure \ref{evolution}, hence it is clear that as in the single mode scheme, long lived $W$ states for systems of the order $n\approx 10$ can be reliably created utilizing this bimodal protocol, under the assumption that the system is in the initial state $|0\rangle$.

\begin{figure*}[top] 
\includegraphics[width=0.9\linewidth]{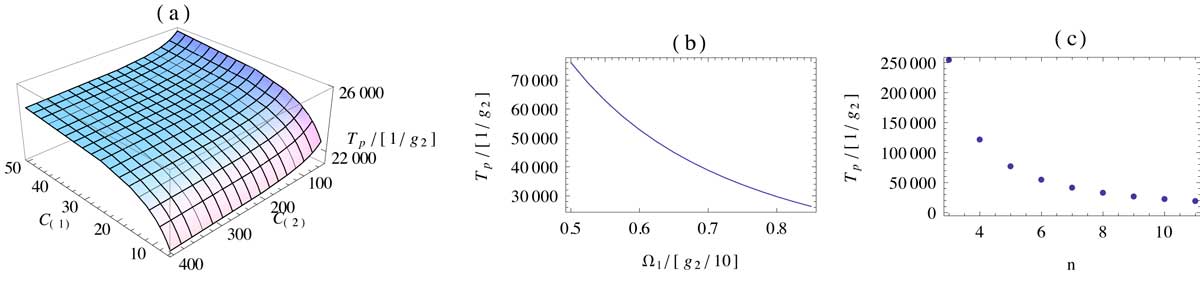} 
\caption{(Color online) (a) Plot of preparation time against cooperativities $C_{(1)}$ and $C_{(2)}$ with a fixed system size of $n=10$. Coherent driving is set within the regime of weak driving with $\Omega_{2} = 1/2$, while $\tilde{\gamma} = 1$ such that cooperativities are varied through $\tilde{\kappa_{1}}$ and $\tilde{\kappa_{2}}$. (b) Plot of preparation time against $\Omega_{2}$ with fixed system size $n=5$. (c) Plot of preparation time against system size with $\Omega_{2} = 1/2$. (b-c) Cooperativities are set with $(\tilde{\gamma},\tilde{\kappa_{1}},\tilde{\kappa_{2}}) = (1,2,1/4)$. (a-c) For all plots $\Omega_{1} = 0$ and    $(\delta_{1},\delta_{2},\Delta_{1},g_{1})$ are at numerically optimized values, as per Eq. \eqref{optimum}, within the restrictions set by Eqs. \eqref{paramch1} and \eqref{paramch2}, necessary for the second phase of the protocol. The threshold value has been set strictly with $z = 0.95$, and the system is assumed to be in the initial state $|3\rangle$. Furthermore, typical values of $g$ are on the order of 10 MHz \cite{kimbleQED}, such that preparation times are on the order of $\mu s$, and $\Omega$ values are on the order of MHz. }\label{BMR} 
\end{figure*}

At this stage it is clear that the bimodal-scheme manages to near perfectly replicate the behavior of the excellent single mode-scheme, assuming the initial state of the system as $|0\rangle$. We now proceed to examine a method for the preparation of the state $|0\rangle$ from an arbitrary initial thermal state of the system, within our bimodal cavity setup.

As seen earlier, $\Omega_{1} = 0$ implies $c_{i}(1) = 0$ such that only effective right to left processes remain. In this case the $n+1$ equations for the diagonal elements of the ground state density matrix, from the effective master equation, again decouple from the equations for the off-diagonal elements. We are left to solve $n+1$ equations of the form,

\begin{equation}
\dot{\rho}_{j} = T_{j}\rho_{j+1} - T_{j-1}\rho_{j},
\end{equation}
where

\begin{equation}
T_{j} = l^{1(b)}_{j} + l^{4(b)}_{j},
\end{equation}
with $l^{1(b)}_{j}$ and $l^{4(b)}_{j}$ defined as,

\begin{align}
& l^{1(b)}_{j} = |\langle j | \hat{L}^{1(b)}_{eff} | j +1 \rangle |^2, \\
& l^{4(b)}_{j} = |\langle j | \hat{L}^{4(b)}_{eff} | j + 1 \rangle |^2.  
\end{align}
The natural steady state of this system is $\rho_{0} = 1$, as desired. However, in order to analyse the efficiency with which this state is created we examine the behavior of the system for one example of some arbitrary initial state. The worst possible case, with $n=3$, corresponds to the initial condition,

\begin{equation}
\rho_{j}(t=0) = \delta_{j,3}.
\end{equation}
We are only interested in the solution corresponding to the state $|0\rangle$, which is given by

\begin{equation}
\rho_{0}(t) = -Ae^{-T_{0}t} - \frac{T_{0}}{T_{1}}B e^{-T_{1}t} 
 - \frac{T_{0}}{T_{2}}C e^{-T_{2}t} + D,
\end{equation}

where the constants are

\begin{align}
&B = -\frac{T_{1}T_{2}}{(T_{0} - T_{1})(T_{1} - T_{2})},  \\
&C = -\frac{T_{1}T_{2}}{(T_{0} - T_{2})(T_{1} - T_{2})},  \\
&A = -(B+C), \\
&D = 1.
\end{align}
It is instantly clear that

\begin{equation}
\lim_{t \rightarrow \infty}\rho_{0}(t) = 1,
\end{equation}
however it still remains to investigate the efficiency with which this limit is obtained, and the dependence of this efficiency on all system parameters. In order to ensure that the W state can be created after the state $|0\rangle$ has been prepared, without altering any cavity parameters, the optimal parameter set \eqref{optimum} will be utilized to describe the detunings. We use $T_{p}$ as a benchmark, where

\begin{equation}
\rho_{0}(T_{p}) = z,
\end{equation}
and $z$ is some threshold accuracy. Figure \ref{BMR} (a) shows that the preparation time has a very weak dependency on $C_{(2)}$, however it exhibits a very favourable dependency on $C_{(1)}$ as it is clear that preparation time actually decreases with decreasing co-operativity. This is excellent, as Figure \ref{BC1} (c) shows that setting $C_{(2)}$ to a very low value will not strongly influence the stability of the W state in the second phase of the scheme.

With respect to $\Omega_{2}$, it is clear from Figure \ref{BMR} (b) that the preparation time of the desired state can be drastically reduced with increased driving strength. As described in \cite{effop}, adiabatic elimination and the effective operator formalism apply only within the regime of weak driving, however all values for $\Omega_{2}$ given in Figure \ref{BMR} (b) fall within a range for which the accuracy of the effective formalism has been well established. 

Finally, Figure \ref{BMR} (c) illustrates the dependency of preparation time on system size. In this case we have that 

\begin{equation}
T_{j} = h(j)\Big[\Omega_{2}^2|\alpha^{(2)}|^2 \big(16\kappa_{1}g_{1}^2 + 2\gamma|\alpha^{(1)}|^2 \big)\Big],
\end{equation}
where we have set $\Delta_{2} = 0$ such that $\alpha^{(k)}_{j} = \alpha^{(k)}$ no longer depends on $j$, and

\begin{equation}
h(j) = \frac{m(j)}{|d_{j}|^2} = \frac{(n-j)(j+1)}{|d_{j}|^2},
\end{equation}
such that due to the form of $h(j)$ increasing system size leads to decreasing preparation times. As can be seen from Figure \ref{BMR} (c) this relationship is approximately exponential, such that although preparation times are significant for smaller systems, for larger systems very reasonable preparation times can be achieved. Again, it is important to note that for all system sizes, increased driving will decrease the preparation times. 

\begin{figure} 
\includegraphics[width=0.9\linewidth]{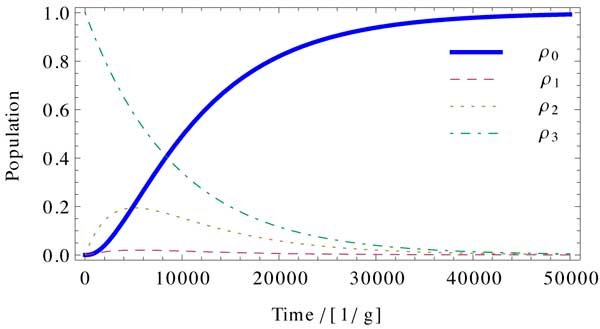} 
\caption{(Color online) Evolution of populations with time for a system of size $n=8$. Coherent driving is set with $\Omega_{1} = 0$ and $\Omega_{2} = 1/2$,  while  $(\delta_{1},\delta_{2},\Delta_{1},g_{1})$ are at numerically optimised values, as per Eq. \eqref{optimum}, within the restrictions set by Eqs. \eqref{paramch1} and \eqref{paramch2}, necessary for the second phase of the protocol. The threshold value has been set strictly with $z = 0.95$, and the system is assumed to be in the initial state $|3\rangle$. Note that typical values of $g$ are on the order of 10 MHz \cite{kimbleQED}, such that preparation times are on the order of $\mu s$. }\label{siml} 
\end{figure}

At this stage it is clear that it is possible to prepare long lived large W states effectively, irrelevant of the initial thermal state of the system. A preliminary step to prepare the state $|0\rangle$ is performed by setting $\Omega_{1} = 0$ and choosing $\Omega_{2}$ appropriately, before the W state is created by setting $\Omega_{2} = 0$ and choosing $\Omega_{1}$ appropriately. In practice this can be done with one laser, whose frequency and strength can be modified after the preparation of $|0\rangle$. Due to the dynamical properties of the system, the state $|0\rangle$ can always be prepared to any accuracy. In practice it is not necessary to prepare the initial state with unity probability, as the right to left nature of the initial scheme is such that for very high populations, all excess population is already in the target state. Finally, it is clear that all benchmarks for the preparation of the W state depend strongly on $C_{(2)}$, but weakly on $C_{(1)}$, while all benchmarks for the preparation of the initial state $|0\rangle$ depend strongly on $C_{(1)}$, but weakly on $C_{(2)}$. This allows co-operativities to be chosen which allow for the optimal performance of both schemes, without having to adjust any cavity parameters between preparation of initial and target state, only the strength and frequency of one laser needs modification.

\section{Conclusions and Outlook}

We have conducted an in-depth study into the construction and properties of the irreducible representations of su(3). From this we have seen that each irreducible representation is invariant under generators of SU(3), such that if one constructs a Hamiltonian and Lindblad operators from these generators, the dynamics of the corresponding open quantum system are constrained to the irreducible subspace to which the initial state belongs. We have applied these properties of SU(3) to ensembles of $\Lambda$ atoms within cavity QED setups, where we have shown that a collective operator approach, embodied through uniform global addressing of atoms within the cavity along with uniform global dissipation, allows one to restrict oneself to a fully symmetric irreducible subspace in which it is possible to apply the effective operator formalism \cite{effop} to arbitrary sized systems. This allows us to expose the effective two-level ground state dynamics for arbitrarily sized systems of $\Lambda$ atoms within an optical cavity.

Furthermore, by application of the above theory we have constructed both a single-mode and bimodal cavity QED setup, in which it is possible to engineer the cavity parameters such that the system is dissipatively driven into a long lived W state for systems on the order of ten atoms, an order of magnitude improvement over all previous schemes of this nature \cite{prep1}-\cite{prepW}. Within the single-mode cavity protocol a specific initial ground state is required, which in practical optical experiments may be easily obtained via Raman pumping, however for the bimodal protocol the target state is obtained irrespective of the initial thermal state of the system. We have performed an in-depth analysis of both protocols, from which it is clear that with currently available optical cavities it is possible to achieve results, comparable with respect to all benchmarks, to previously suggested protocols for three atoms \cite{prepW}. Furthermore, the protocols suggested here require only one laser, for uniform global addressing of all atoms, a vast improvement over previously suggested schemes.

Importantly the characteristic behaviour of both protocols, with respect to all relevant benchmarks, displays excellent scaling properties against cavity cooperativity. This indicates that with inevitable experimental developments and the availability of high cooperativity cavities, it should soon be possible to implement these schemes and obtain extremely long lived W states, with excellent fidelity, for large systems on the order of ten atoms. Furthermore, for general systems the witness methods of \cite{WstateWitness} offer a possible means for state characterization, while for QED setups, as discussed in this paper, atomic state tomography \cite{tomography} may be utilized for the characterization and verification of these results.

\begin{acknowledgments}
This work is based upon research supported by the South African
Research Chair Initiative of the Department of Science and
Technology and National Research Foundation. R.S. also acknowledges support from the National Institute for Theoretical Physics.

\end{acknowledgments}

\begin{appendix}

\section{Properties of SU(3)}

The full set of generators, $\{\hat{\lambda}_{i}\}$, for SU(3) are as follows: The first three are constructed by an extension of the Pauli Matrices into an extra dimension,

\[
\hat{\lambda}_{1} =
 \begin{pmatrix}
0&1&0\\
1&0&0\\
0&0&0
\end{pmatrix} \qquad
\hat{\lambda}_{2} =
 \begin{pmatrix}
0&-i&0\\
i&0&0\\
0&0&0
\end{pmatrix}
\]

\[
\hat{\lambda}_{3} =
 \begin{pmatrix}
1&0&0\\
0&-1&0\\
0&0&0
\end{pmatrix} 
\]

These are clearly traceless and Hermitian by construction. The remaining 5 generators have been chosen as per the conventions in particle physics, in clear analogy with the Pauli matrices,

\[
\hat{\lambda}_{4} =
 \begin{pmatrix}
0&0&1\\
0&0&0\\
1&0&0
\end{pmatrix} \qquad
\hat{\lambda}_{5} =
 \begin{pmatrix}
0&0&-i\\
0&0&0\\
i&0&0
\end{pmatrix}
\]

\[
\hat{\lambda}_{6} =
 \begin{pmatrix}
0&0&0\\
0&0&1\\
0&1&0
\end{pmatrix} \qquad
\hat{\lambda}_{7} =
 \begin{pmatrix}
0&0&0\\
0&0&-i\\
0&i&0
\end{pmatrix}
\]

\[
\hat{\lambda}_{8} = \frac{1}{\sqrt{3}}
 \begin{pmatrix}
1&0&0\\
0&1&0\\
0&0&-2
\end{pmatrix} 
\]

The commutators of the above generators are given by 

\begin{equation}
[\hat{\lambda}_{i},\hat{\lambda}_{j}] = 2 i f_{ijk}\hat{\lambda}_{k}
\end{equation}

where the structure constants are totally antisymmetric under exchange of any two indices, and their non-vanishing values are listed below in Table \ref{tab1}.

\begin{table}[H]
\begin{center}
\caption{Non-vanishing structure constants, $\{f_{ijk}\}$, up to antisymmetric permutations.}\label{tab1}
\begin{tabular}{ c|ccccccccc  }

  $ijk$ & 123 & 147  &  156 &  246  & 257  & 345  & 367  &  458  &  678\\ [2pt] \hline
 $f_{ijk}$&  1  &  $\frac{1}{2}$  & $-\frac{1}{2}$  & $\frac{1}{2}$  &  $\frac{1}{2}$  &  $\frac{1}{2}$  & - $\frac{1}{2}$  &$\frac{\sqrt{3}}{2}$   & $\frac{\sqrt{3}}{2}$  \\
\end{tabular}

\end{center}
\end{table}

For the purposes of this paper, it convenient to construct operators $\hat{T}_{\pm}$, $\hat{V}_{\pm}$, $\hat{U}_{\pm}$, $\hat{T}_{3}$, $\hat{Y}$ as per equations \eqref{Fspin1}, \eqref{Fspin2}, \eqref{Fspin3}. The full set of commutation relationships for these operators is as follows, 

\begin{align}
&[\hat{T}_{3},\hat{T}_{\pm}] = \pm\hat{T}_{\pm} \qquad [\hat{T}_{+},\hat{T}_{-}] = 2\hat{T}_{3} \label{commute1}\\
&[\hat{T}_{3},\hat{U}_{\pm}] = \mp\frac{1}{2}\hat{U}_{\pm} \quad [\hat{U}_{+},\hat{U}_{-}] = \frac{3}{2}\hat{Y} - \hat{T}_{3} \equiv 2\hat{U}_{3} \label{commute2_}\\
&[\hat{T}_{3},\hat{V}_{\pm}] = \pm\frac{1}{2}\hat{V}_{\pm} \quad [\hat{V}_{+},\hat{V}_{-}] = \frac{3}{2}\hat{Y} + \hat{T}_{3} \equiv 2\hat{V}_{3}\label{commute3}\\
&[\hat{Y},\hat{T}_{\pm}] = 0 \quad [\hat{Y},\hat{U}_{\pm}] = \pm\hat{U}_{\pm} \quad [\hat{Y},\hat{V}_{\pm}] = \pm\hat{V}_{\pm}\label{commute4}\\
&[\hat{T}_{+},\hat{V}_{+}] = [\hat{T}_{+},\hat{U}_{-}] = [\hat{U}_{+},\hat{V}_{+}] = 0 \label{commute5}\\
&[\hat{T}_{+},\hat{V}_{-}] = -\hat{U}_{-} \quad [\hat{T}_{+},\hat{U}_{+}] = \hat{V}_{+} \label{commute6}\\
&[\hat{U}_{+},\hat{V}_{-}] = \hat{T}_{-} \quad [\hat{T}_{3},\hat{Y}] = 0 \label{commute7}
\end{align}

\end{appendix}

\end{document}